\documentclass[aps,prd,nofootinbib,superscriptaddress,preprint]{revtex4-1}
\pdfoutput=1
\usepackage[utf8]{inputenc}
\usepackage{amsmath,amssymb}
\usepackage{epsfig}
\usepackage{graphicx}
\usepackage{slashed}
\usepackage[usenames,dvipsnames]{color}
\usepackage[colorlinks,citecolor=blue]{hyperref}
\usepackage{soul}
\usepackage{color}
\usepackage{subfigure}
\begin{document}

\title{Cosmic Inflation in Minimal $U(1)_{B-L}$ Model: Implications for (Non) Thermal Dark Matter and Leptogenesis}

\author{Debasish Borah}
\email{dborah@iitg.ac.in}
\affiliation{Department of Physics, Indian Institute of Technology Guwahati, Assam 781039, India}
\author{Suruj Jyoti Das }
\email{suruj@iitg.ac.in}
\affiliation{Department of Physics, Indian Institute of Technology Guwahati, Assam 781039, India}
\author{Abhijit Kumar Saha}
\email{aks@prl.res.in}
\affiliation{Theoretical Physics Division, Physical Research Laboratory, Navrangpura, Ahmedabad 380009, India}
\footnote{During the review process of this manuscript, the affiliation of AKS has been changed to “School of Physical Sciences, Indian Association for the Cultivation of Science, 2A and 2B Raja S.C. Mullick Road, Kolkata 700032, India”. The new institutional email address of AKS is psaks2484@iacs.res.in.}

\begin{abstract}
We study the possibility of realising cosmic inflation, dark matter (DM), baryon asymmetry of the universe (BAU) and light neutrino masses in non-supersymmetric minimal gauged $B-L$ extension of the standard model with three right handed neutrinos. The singlet scalar field responsible for spontaneous breaking of $B-L$ gauge symmetry also plays the role of inflaton by virtue of its non-minimal coupling to gravity. While the lightest right handed neutrino is the DM candidate, being stabilised by an additional $Z_2$ symmetry, we show by performing a detailed renormalisation group evolution (RGE) improved study of inflationary dynamics that thermal DM is generally overproduced due to insufficient annihilations through gauge and scalar portals. This happens due to strict upper limits obtained on gauge and other dimensionless couplings responsible for DM annihilation while assuming the non-minimal coupling to gravity to be at most of order unity. The non-thermal DM scenario is viable, with or without $Z_2$ symmetry, although in such a case the $B-L$ gauge sector remains decoupled from the inflationary dynamics due to tiny couplings. We also show that the reheat temperature predicted by the model prefers non-thermal leptogenesis with hierarchical right handed neutrinos while being consistent with other requirements.
\end{abstract}

\maketitle

\section{Introduction}
Precision measurements of the cosmic microwave background (CMB) anisotropies by experiments like Planck \cite{Komatsu:2010fb, Akrami:2018odb, Aghanim:2018eyx} reveal that our universe is homogeneous and isotropic on large scales upto a remarkable accuracy. However, the observed isotropy of the CMB leads to the horizon problem which remains unexplained in the standard cosmology where the universe remains radiation dominated throughout the early stages. In order to solve the horizon problem, the presence of a rapid accelerated expansion phase in the early universe, called inflation \cite{Guth:1980zm,Starobinsky:1980te, Linde:1981mu} was proposed. Originally proposed to solve the horizon, flatness and unwanted relic problems in standard cosmology, the inflationary paradigm was also subsequently supported by the adiabatic and scale invariant perturbations observed in the CMB \cite{Komatsu:2010fb, Akrami:2018odb}. Such an early accelerated phase of expansion can be generated by the presence of one or more scalar fields whose dynamics crucially decides the period of inflation. Over the years, a variety of inflationary models have been studied with different levels of success~\cite{Mazumdar:2010sa}. The earliest proposal of this sort is known as chaotic inflation \cite{Linde:1983gd, Martin:2013tda} where simple power law potentials like $m^2 \phi^2$ with a scalar field $\phi$ were used. However, such simple models predict very specific values of inflationary parameters like the spectral index $n_s\sim 0.967$, tensor-to-scalar ratio $r\sim 0.133$ for number of e-folds $N_e=60$ and unfortunately, the latest Planck 2018 data \cite{Akrami:2018odb} strongly disfavour this simple model due to its large prediction of $r$. Modified chaotic inflation where the inflation sector is extended by an additional scalar field to assist the inflaton field has also been proposed  \cite{Harigaya:2015pea,Saha:2016ozn,Borah:2019bdi}. Another class of models use the Higgs as the inflaton \cite{Bezrukov:2007ep, Bezrukov:2010jz}. These models often suffer from problems of vacuum stability \cite{Sher:1988mj} and non-unitarity \cite{Lerner:2009na} as well as being inadequate for combining inflation with other cosmological problems like DM and BAU. A possible way out is to consider a beyond standard model (BSM) singlet scalar which acts as the inflaton. We consider this possibility in our work where an additional scalar with non-minimal coupling to gravity \cite{Pallis:2014cda,Gumjudpai:2016ioy,Tenkanen:2017jih, Shokri:2019rfi}, in addition to usual quartic chaotic type coupling, can give rise to successful inflation while predicting the inflationary parameters within the observed range. The same scalar field is also responsible for several other interesting phenomenology as we discuss below.

The same CMB measurements mentioned above also suggest that the present universe has a significant amount of non-luminous, non-baryonic form of matter, known as dark matter (DM) \cite{Aghanim:2018eyx, Tanabashi:2018oca}. This is also supported by astrophysical evidences gathered over a much longer period of time \cite{Zwicky:1933gu, Rubin:1970zza, Clowe:2006eq}. The Planck 2018 data reveals that approximately $26\%$ of the present universe is composed of DM, which is about five times more than the ordinary luminous or baryonic matter. In terms of density parameter $\Omega_{\rm DM}$ and $h = \text{Hubble Parameter}/(100 \;\text{km} ~\text{s}^{-1} \text{Mpc}^{-1})$, the present DM abundance is conventionally reported as \cite{Aghanim:2018eyx}:
$\Omega_{\text{DM}} h^2 = 0.120\pm 0.001$
at 68\% CL. Since none of the standard model (SM) particles can satisfy the criteria of a particle DM candidate, several proposals have been put forward among which the weakly interacting massive particle (WIMP) is perhaps the most popular one. In this framework, a DM particle having mass and interactions typically around the electroweak scale can give rise to the observed DM abundance after thermal freeze-out, a remarkable coincidence often referred to as the {\it WIMP Miracle} \cite{Kolb:1990vq}. The same interactions responsible for thermal freeze-out of WIMP type DM should also give rise to sizeable DM-nucleon scattering. However, null results at direct detection experiments like LUX \cite{Akerib:2016vxi}, PandaX-II \cite{Tan:2016zwf, Cui:2017nnn}, XENON1T \cite{Aprile:2017iyp, Aprile:2018dbl} have certainly pushed several WIMP models into a tight corner, if not ruled out yet. This has also generated interests in beyond thermal WIMP paradigms as viable alternatives. One such interesting possibility is the non-thermal origin of DM \cite{Hall:2009bx}. For a recent review of such feebly interacting (or freeze-in) massive particle (FIMP) DM, please see \cite{Bernal:2017kxu}. In the FIMP scenario, DM candidate does not thermalise with the SM particles in the
early universe due to its feeble interaction strength and the initial abundance of DM
is assumed to be zero. At some later stage, DM can be produced non thermally from decay or annihilation of other particles thermally present in the universe.

Similarly, the baryonic content of the universe also gives rise to another puzzle due to the abundance of baryons over antibaryons. Quantitatively, this excess is denoted as baryon to entropy ratio \cite{Tanabashi:2018oca, Aghanim:2018eyx}
\begin{equation}
Y_B = \frac{n_{B}-n_{\bar{B}}}{s} = (8.24 -9.38) \times 10^{-10}
\label{etaBobs}
\end{equation}
where $Y_B$ denotes comoving baryon density, $n_B (n_{\bar{B}})$ denotes baryon (anti-baryon) number density while $s$ is the entropy density. Since any initial asymmetry before inflation will be washed out at the end of inflation due to the exponential expansion of the universe, there has to be a dynamical mechanism to generate the asymmetry in a post-inflationary universe. This requires certain conditions, known as the Sakharov conditions~\cite{Sakharov:1967dj} to be fulfilled. They are namely, baryon number (B) violation, C and CP violation and departure from thermal equilibrium, not all of which can be fulfilled in the required amounts within the SM alone. Generation of baryon asymmetry of the universe (BAU) from out-of-equilibrium decays of heavy particles has been a well-known mechanism for baryogenesis~\cite{Weinberg:1979bt, Kolb:1979qa}. Another interesting way, which also connects the lepton sector physics, is known as leptogenesis, proposed a few decades back \cite{Fukugita:1986hr}. In leptogenesis, instead of creating a baryon asymmetry directly from B violating interactions, an asymmetry in lepton sector is created via lepton number (L) violating processes (decay or scattering). If this lepton asymmetry is generated before the electroweak phase transition (EWPT), then the $(B+L)$-violating electroweak sphaleron transitions~\cite{Kuzmin:1985mm} can convert it to the required baryon asymmetry. Since the quark sector CP violation is insufficient to produce the required baryon asymmetry, the mechanism of leptogenesis can rely upon lepton sector CP violation which may be quite large as hinted by some neutrino oscillation experiments \cite{Esteban:2018azc, Abe:2019vii}. An interesting feature of this scenario is that the required lepton asymmetry can be generated through CP violating out-of-equilibrium decays of the same heavy fields that take part in popular seesaw mechanisms~\cite{Minkowski:1977sc, Mohapatra:1979ia, Yanagida:1979as, GellMann:1980vs, Glashow:1979nm, Schechter:1980gr} which also explains the origin of tiny neutrino masses~\cite{Tanabashi:2018oca}, another observed phenomena which the SM fails to address.

Motivated by these, we study a minimal extension of the SM, by a gauged $B-L$ symmetry with three right handed neutrinos (RHN) required to cancel the anomalies and a singlet scalar to break the additional gauge symmetry spontaneously while simultaneously generating RHN masses. Although previously analysed separately, the consistency of these three entities together have not been examined in this simple kind of BSM setup before as per our knowledge. We also perform a complete RG evolution of all the relevant couplings to determine the fate of the scenarios we discuss here. While in this framework, the singlet scalar plays the role of inflaton, one RHN is stabilised by an additional $Z_2$ symmetry to become a DM candidate. The other two RHNs can give rise to light neutrino masses with vanishing lightest neutrino mass apart from producing the required lepton asymmetry which gets converted into the observed baryon asymmetry via sphalerons. Interestingly, we find that the stringent limits on the inflationary observables from Planck 2018 and BICEP 2 / Keck Array (BK15) data \cite{Akrami:2018odb} as well as the stability of inflaton potential restrict the $B-L$ gauge coupling, scalar couplings and Yukawa couplings associated with the inflaton field to be within some limits which do not favour thermal DM scenario due to insufficient annihilations. As an alternative, with very tiny gauge and Yukawa couplings, one can realise the non-thermal DM scenario (with or without $Z_2$ symmetry) while the inflationary potential behaviour merges with the usual case of quartic inflation with non minimal coupling to gravity. We also find that the predicted values of reheat temperature makes it difficult to realise high scale thermal $N_2$ leptogenesis \cite{DiBari:2005st, Mahanta:2019gfe} with hierarchical RHN leaving the option of non-thermal leptogenesis \cite{Lazarides:1991wu, Giudice:1999fb, Asaka:1999yd, Asaka:1999jb, Fujii:2002jw, Pascoli:2003rq, Asaka:2002zu, Panotopoulos:2006wj, HahnWoernle:2008pq} viable.

The structure of the paper is organised as follows. In section \ref{sec:model}, we discuss the particle content of the proposed setup and their interactions followed by brief mention of the existing constraints in section \ref{sec:const}. In section \ref{sec:inf} we perform a detailed study of inflation and its predictions in view of Planck 2018 bounds. We discuss different aspects of DM phenomenology in section \ref{sec:DM} and then move onto discussing the possibility of non-thermal leptogenesis in section \ref{sec:baryo}. Finally we conclude in section \ref{sec:conclude}.

\section{The Model}\label{sec:model}
As mentioned earlier, we study a gauged $B-L$ extension of the SM with the minimal field content which can give rise to cancellation of triangle anomalies, spontaneous gauge symmetry breaking, light neutrino masses, dark matter, leptogenesis and cosmic inflation. While gauged $B-L$ extension of the SM was proposed long ago \cite{Davidson:1978pm, Mohapatra:1980qe, Marshak:1979fm, Masiero:1982fi, Mohapatra:1982xz, Buchmuller:1991ce}, realising a stable DM candidate in the model requires non-minimal field content or additional discrete symmetries. Also, a gauged $B-L$ model with just SM fermion content, is not anomaly free due to the non-vanishing triangle anomalies for both $U(1)^3_{B-L}$ and mixed $U(1)_{B-L}-(\text{gravity})^2$ anomalies. These triangle anomalies for the SM fermion content are given as
\begin{align}
\mathcal{A}_1 \left[ U(1)^3_{B-L} \right] = \mathcal{A}^{\text{SM}}_1 \left[ U(1)^3_{B-L} \right]=-3\,,  \nonumber \\
\mathcal{A}_2 \left[(\text{gravity})^2 \times U(1)_{B-L} \right] = \mathcal{A}^{\text{SM}}_2 \left[ (\text{gravity})^2 \times U(1)_{B-L} \right]=-3\,.
\end{align}
Remarkably, if three right handed neutrinos with $B-L$ charge -1 each are added to the model, they contribute $\mathcal{A}^{\text{New}}_1 \left[ U(1)^3_{B-L} \right] = 3, \mathcal{A}^{\text{New}}_2 \left[ (\text{gravity})^2 \times U(1)_{B-L} \right] = 3$ leading to vanishing amount of triangle anomalies. This is perhaps the most economical setup of anomaly cancellation and hence we adopt it here \footnote{For other exotic solutions to anomaly cancellation conditions, see \cite{Montero:2007cd, Wang:2015saa, Patra:2016ofq, Nanda:2017bmi, Bernal:2018aon, Biswas:2019ygr, Nanda:2019nqy}.}. To have a stable DM candidate we introduce a discrete $Z_2$ symmetry under which one of the RHN is odd whereas all other fields are even. In Tables \ref{tab:1} and \ref{tab:2}, we have listed all fermions as well as scalar fields (including the SM ones) of the present model and their charges under the $SU(3)_c \times SU(2)_L \times U(1)_Y \times U(1)_{B-L}$ symmetry.   

\begin{table}
\begin{center}
\begin{tabular}{|c|c|c|}
\hline
Particles & $SU(3)_c \times SU(2)_L \times U(1)_Y \times U(1)_{B-L} $ & $Z_2$  \\
\hline
$q_L=\begin{pmatrix}u_{L}\\
d_{L}\end{pmatrix}$ & $(3, 2, \frac{1}{6}, \frac{1}{3})$ & +  \\
$u_R$ & $(3, 1, \frac{2}{3}, \frac{1}{3})$ & +  \\
$d_R$ & $(3, 1, -\frac{1}{3}, \frac{1}{3})$ & +  \\

$\ell_L=\begin{pmatrix}\nu_{L}\\
e_{L}\end{pmatrix}$ & $(1, 2, -\frac{1}{2}, -1)$ & +  \\
$e_R$ & $(1, 1, -1, -1)$ & +\\
\hline
$N_{R_1}$ & $(1, 1, 0, -1)$ & - \\
$N_{R_2}$ & $(1, 1, 0, -1)$ & + \\
$N_{R_3}$ & $(1, 1, 0, -1)$ & + \\
\hline
\end{tabular}
\end{center}
\caption{Fermion fields of the model and their corresponding gauge charges.}
\label{tab:1}
\end{table}
\begin{table}
\begin{center}
\begin{tabular}{|c|c|c|}
\hline
Particles & $SU(3)_c \times SU(2)_L \times U(1)_Y \times U(1)_{B-L} $  & $Z_2$  \\
\hline
$H=\begin{pmatrix}H^+\\
H^0\end{pmatrix}$ & $(1,2,\frac{1}{2},0)$ & +  \\
\hline
$\Phi$ & $(1, 1, 0, 2)$ & + \\
\hline
\end{tabular}
\end{center}
\caption{Scalar fields of the model and their corresponding gauge charges.}
\label{tab:2}
\end{table}
The gauge invariant Lagrangian of the model is 
\begin{eqnarray}
\mathcal{L}&=&\mathcal{L}_{\rm SM} -\frac{1}{4} {B^{\prime}}_{\alpha \beta}
\,{B^{\prime}}^{\alpha \beta} + \mathcal{L}_{\rm scalar} 
+ \mathcal{L}_{\rm fermion}\;.
\label{LagT}
\end{eqnarray}
where $\mathcal{L}_{\rm SM}$ denotes the SM Lagrangian involving quarks,
gluons, charged leptons, left handed neutrinos and electroweak gauge
bosons while the second term is the kinetic term of $B-L$ gauge boson ($Z_{BL}$)
expressed in terms of field strength tensor ${B^\prime}^{\alpha\beta}=
\partial^{\alpha}Z_{BL}^{\beta}-\partial^{\beta}Z_{BL}^{\alpha}$. The gauge invariant scalar Lagrangian of the model is as follows
\begin{align}
 \mathcal{L}_{\rm scalar}=(D_{\mu} H)(D_{\mu} H)^\dagger+(D_{\mu} \Phi)(D^{\mu} \Phi)^\dagger-V(H,\Phi)~,
\end{align}
where
\begin{align}
V(H,\Phi)=-\mu_{1}^{2} |H|^{2}-\mu_{2}^{2}|\Phi|^{2}+\lambda_{1} |H|^{4}+\lambda_{2}|\Phi|^{4}+\lambda_{3} |H|^{2}|\Phi|^{2}.\label{eq:PotI}
\end{align}
The covariant derivatives of scalar fields are
\begin{align}
 &D_{\mu} H=\left(\partial_\mu+i\frac{g_1}{2}\sigma_aW^a_{\mu}+i\frac{g_2}{2}B_\mu\right)H,\\
  &D_{\mu} \Phi =\left(\partial_\mu+i2g_{BL}Z_{BL\mu}\right)\Phi,
\end{align}
with $g_1$ and $g_2$ being the gauge couplings of $SU(2)_L$ and $U(1)_Y$ respectively
and $W^a_{\mu}$
($a=1,\,2,\,3$) and $B_{\mu}$ are the corresponding gauge fields. On the other hand $Z_{BL}, g_{BL}$ are the gauge boson and gauge coupling respectively for $U(1)_{B-L}$ gauge group.

The gauge invariant fermionic Lagrangian of the model is as follows
\begin{align}
 \mathcal{L}_{\rm fermion} &=  i \sum_{\kappa=1}^{3}\overline{N_{R_{\kappa}}} \slashed{D}(Q^R_{\kappa}) N_{R_{\kappa}} -\sum_{j=2}^{3}\sum_{\alpha=e, \mu, \tau} Y_{D}^{j \alpha}~\overline{l_{L}^{\alpha}}\tilde{H}N_{R}^{j}-\sum_{i=2}^{3}\sum_{j=2}^{3}Y_{N_{ij}}\Phi~\overline{N_{R_i}^{C}}N_{R_j} \nonumber \\
& -Y_{N_1}\Phi~\overline{N_{R_1}^{C}}N_{R_1}+{\rm h.c.}\label{eq:Lferm}
\end{align}
The covariant derivative is defined as 
\begin{eqnarray}
\slashed{D}(Q^{R}_{\kappa})\,{N_{R_{\kappa}}} =
\gamma^{\mu}\left(\partial_{\mu}
+ i g_{BL}\,Q^{(R)}_{\kappa}\,{Z_{BL}}_{\mu}\right) {N_{R_{\kappa}}} \,, 
\end{eqnarray}
with $Q^R_{\kappa}=-1$ being the $B-L$ charge of right handed neutrino $N_{R_{\kappa}}$. Due to the presence of $Z_2$ symmetry, $N_{R_1}$ has no mixing with $N_{R_{2,3}}$ and also does not interact with SM leptons thereby qualifying for a stable DM candidate.

After breaking of both $B-L$ symmetry and electroweak symmetry
by the vacuum expectation values (VEVs) of $H$ and $\Phi$, the form of doublet and singlet scalar fields
are given by,
\begin{eqnarray}
H=\begin{pmatrix}H^+\\
\dfrac{h + v + i A}{\sqrt{2}}\end{pmatrix}\,,\,\,\,\,\,\,
\Phi = \dfrac{\phi+v_{BL}+ iA^{\prime}}{\sqrt{2}}
\label{H&phi_broken_phsae}
\end{eqnarray}
where $v$ and $v_{BL}$ are VEVs of $H$
and $\Phi$ respectively. The right handed neutrinos and $Z_{BL}$ get masses after the $U(1)_{B-L}$ breaking as, 
\begin{align}
 &M_{Z_{BL}}=2 g_{BL} v_{BL,}\label{eq:Zmass}\\
 &M_{N_{i}}=\sqrt{2}Y_{N_{i}}v_{BL}.\label{eq:DMmass}
\end{align}
Here we consider diagonal Yukawa $Y_N$ in $(N_{R_1},N_{R_2}, N_{R_3})$ basis. Using equation (\ref{eq:Zmass}) and equation (\ref{eq:DMmass}), it is possible to relate  $M_{Z_{BL}}$ and $M_{N_i}$ by,
\begin{align}
 M_{N_{i}}=\frac{1}{\sqrt{2} g_{BL}}Y_{N_{i}} M_{Z_{BL}}.
\end{align}
Also after the breaking of $SU(2)_{L}\times U(1)_{Y}\times U(1)_{B-L}$, the scalar fields $h$ and $\phi$ can be related to the physical mass eigenstates $H_{1}$ and $H_{2}$ by a rotation matrix as, 
\begin{align}
\begin{pmatrix}H_{1}\\
H_{2}
\end{pmatrix}=\begin{pmatrix}\cos\theta & -\sin\theta\\
\sin\theta & \cos\theta
\end{pmatrix}\begin{pmatrix}h\\
\phi
\end{pmatrix},
\end{align}
where the scalar mixing angle $\theta$ is represented by 
\begin{align}
\tan2\theta=-\frac{\lambda_{3}vv_{BL}}{(\lambda_{1}v^2-\lambda_{2}v_{BL}^2)}~.
\label{mixingangle}
\end{align}  
The physical scalar masses are given by, 
\begin{align}
 &M_{H_{1}}^2=2{\lambda_{1}}v^2\cos^2\theta+2{\lambda_{2}}v_{BL}^2\sin^2\theta-2{\lambda_{3}}vv_{BL}\sin\theta\cos\theta,\\
 &M_{H_{2}}^2=2{\lambda_{1}}v^2\sin^2\theta+2{\lambda_{2}}v_{BL}^2\cos^2\theta+2{\lambda_{3}}vv_{BL}\sin\theta\cos\theta.\label{eq:HiggsEigen}
\end{align}
Here $M_{H_1}$ is identified as the SM Higgs mass whereas $M_{H_2}$ is the singlet scalar mass.

One of the strong motivations of the minimal $U(1)_{B-L}$ model is the presence of heavy RHNs which can yield correct light neutrino mass via type I seesaw mechanism. The analytical expression for the light neutrino mass matrix is
\begin{align}
 m_{\nu}=m_D^TM_{N}^{-1}m_D,
\end{align}
where $m_D=Y_Dv/\sqrt{2}$. We consider the right handed neutrino mass matrix $M_N$ to be diagonal. Since in our case $N_{R_1}$ does not interact with SM leptons, the lightest active neutrino would be massless.
The Dirac neutrino Yukawa matrix $Y_D$ can be formulated through the Casas-Ibarra parametrisation \cite{Casas:2001sr} as  
\begin{align}\label{casas}
    Y_D=\sqrt{2}\frac{\sqrt{M_N}}{v}  \mathcal{R}\sqrt{m_\nu^d} ~U^\dagger_{\textrm{PMNS}},
    \end{align}
where $m_\nu^d, M_N$ are the diagonal light and heavy neutrino mass matrices respectively and $U_{\textrm{PMNS}}$ is the usual Pontecorvo-Maki-Nakagawa-Sakata (PMNS) leptonic mixing matrix. In the diagonal charged lepton basis, the PMNS mixing matrix is also the diagonalising matrix of light neutrino mass matrix  
$$m_{\nu} = U^*_{\textrm{PMNS}} m_\nu^d U^\dagger_{\textrm{PMNS}}.$$
In the above Casas-Ibarra parametrisation, $\mathcal{R}$ represents a complex orthogonal matrix ($\mathcal{R}\mathcal{R}^{T}=\mathcal{I}$). In case of only two right handed neutrinos, the $\mathcal{R}$ matrix is a function of only one complex rotation parameter $z=z_R + i z_I, z_R \in [0, 2\pi], z_I \in \mathbb{R}$ \cite{Ibarra:2003up}. For three right handed neutrinos taking part in seesaw mechanism $\mathcal{R}$ can depend upon three complex rotation parameters. Assuming one of them (rotation in 1-2 sector) to be vanishing, it can be represented as\footnote{For some recent discussions on choice of $\mathcal{R}$ matrix in the context of thermal and non-thermal dark matter as well as leptogenesis, please see \cite{Mahanta:2019gfe}.}
\begin{align}
 \mathcal{R}=\begin{pmatrix}
              \cos \gamma^\prime & 0 &\sin \gamma^\prime\\
              -\sin \gamma\sin \gamma^\prime & \cos \gamma & \sin \gamma\cos \gamma^\prime\\
              -\cos \gamma\sin \gamma^\prime & -\sin \gamma & \cos \gamma\cos \gamma^\prime
             \end{pmatrix}.
             \label{Rmatrix}
\end{align}
Therefore with suitable choices of $\gamma$ and $\gamma^\prime$, the Yukawa matrix can take different forms. Here it remains pertinent to note that for a $Z_2$ symmetric Lagrangian ($\gamma^\prime \sim 0$) as described in equation (\ref{eq:Lferm}), the Dirac Yukawa coupling $Y_D$ represents a $2\times 3$ matrix in flavour basis. We shall use the best fit values of all three mixing angles and the mass squared differences of active neutrinos assuming a normal ordering \cite{Tanabashi:2018oca}.

\section{Constraint on the Model Parameters}
\label{sec:const}
In this section, we briefly discuss the theoretical and experimental constraints on different parameters of the model. 

To begin with, we consider the bounded from below criteria of the scalar potential. This gives rise to the following conditions to be satisfied by the quartic couplings,
$$ \lambda_{1, 2, 3} \geq 0, \, \lambda_3 + \sqrt{\lambda_1 \lambda_2} \geq 0 $$.
On the other hand, to avoid perturbative breakdown of the model, all dimensionless couplings must obey the following limits at any energy scale:
$$ | \lambda_{1,2,3} | < 4\pi, \, |Y_D, Y_N| < \sqrt{4\pi}, \, |g_1, g_2, g_{BL} | < \sqrt{4\pi}. $$

The non-observation of the extra neutral gauge boson in the LEP experiment \cite{Carena:2004xs,Cacciapaglia:2006pk} invokes following constraint on the ratio of $M_{Z_{BL}}$ and $g_{BL}$ :
\begin{align}
 \frac{M_{Z_{BL}}}{g_{BL}} \geq 7 {~\rm TeV}.
\end{align}
The corresponding bounds from the large hadron collider (LHC) experiment have become stronger than this by now as both the ATLAS and the CMS collaborations have performed dedicated searches for dilepton resonances in proton-proton collisions. The latest bounds from the ATLAS experiment \cite{Aaboud:2017buh, Aad:2019fac} and the CMS experiment \cite{Sirunyan:2018exx} at the LHC rule out such gauge boson masses below 4-5 TeV from analysis of 13 TeV centre of mass energy data. However, such limits are derived by considering the corresponding gauge coupling $g_{BL}$ to be similar to the ones in electroweak theory and hence the bounds become less stringent for weaker gauge couplings \cite{Aaboud:2017buh}. Additionally, if such Abelian gauge bosons couple only to the third generation leptons, then the collider bounds get even weaker, as explored recently in a singlet-doublet fermion DM scenario by the authors of \cite{Barman:2019aku}.

Additionally, the singlet scalar of the model is also constrained \cite{Robens:2015gla,Chalons:2016jeu} as it can mix with the SM Higgs and hence can couple to SM fields. The strongest bound on such mixing in scalar singlet extension of the SM arises from $W$ boson mass correction \cite{Lopez-Val:2014jva} at NLO. For singlet scalar mass $250 {\rm ~ GeV} \lesssim M_{H_2} \lesssim 850$ GeV, the singlet-SM Higgs mixing is constrained to be $0.2 \lesssim \sin\theta \lesssim 0.3$. For heavier singlet scalar masses $M_{H_2} > 850$ GeV, the bounds from the requirement of perturbativity and unitarity of the theory turn dominant  which gives $\sin\theta \lesssim 0.2$. On the other hand, for lighter singlet scalar masses $M_{s_i}<250$ GeV, the LHC and LEP direct search \cite{Khachatryan:2015cwa,Strassler:2006ri} and Higgs signal strength measurement  \cite{Strassler:2006ri} constrain the mixing angle as $\sin{\theta} \lesssim 0.25$. If the singlet scalar is even lighter say, lighter than SM Higgs mass $M_{H_2} < M_{H_1}/2$, SM Higgs can decay into a pair of singlet scalars. Latest measurements by the ATLAS collaboration restrict such SM Higgs decay branching ratio into invisible particles to be below $13\%$ \cite{ATLAS:2020cjb} at $95\%$ CL.

\section{Inflation}\label{sec:inf}
In this section, we describe the dynamics of inflation in detail and its predictions in view of the present experimental bounds. We identify the real part of singlet scalar field $\Phi$ as the inflaton. Along with the renormalisable potential in equation (\ref{eq:PotI}), we also assume that $\Phi$ is non-minimally coupled to gravity. For earlier studies in this context, please see \cite{Okada:2011en, Okada:2015lia} and references therein. Related studies in supersymmetric gauged $B-L$ model can be found in \cite{Buchmuller:2012wn}. For works guided by the same unifying principle of inflation, dark matter and neutrino mass, one may look at \cite{Allahverdi:2007wt, Kazanas:2004kv, Dong:2018aak, Borah:2018rca} as well as references therein.

We denote the inflaton field as $\phi$ hereafter, which is same as the notation used for real part of $\Phi$ field in earlier sections. Thus the potential responsible for inflation is given by
\begin{align}
 V_{\rm Inf}(\phi)=\frac{\lambda_2}{4} \phi^4+ \frac{\xi}{2} \phi^2 R,\label{eq:PotPhi}
\end{align}
where $R$ stands for the Ricci scalar and $\xi$ is a dimensionless coupling of singlet scalar to gravity. We have neglected the contribution of $v_{BL}$ in equation (\ref{eq:PotPhi}) by considering it to be much lower than the reduced Planck mass $M_P$. The action for $\phi$ in Jordan frame takes the following form (apart from the couplings to the fermions and SM Higgs)
\begin{align}
 S_J=\int d^4x\sqrt{-g}\Bigg[-\frac{M_P^2}{2}\Omega(\phi)^2R+\frac{1}{2}(D_{\mu}\phi)^\dagger(D^{\mu}\phi)-\frac{\lambda_2}{4} \phi^4\Bigg],
\end{align}
where $\Omega(\phi)^2=1+ \frac{\xi\phi^2}{M_P^2}$, $g$ is the spacetime metric in the $(-,+,+,+)$ convention, $D_{\mu} \phi$ stands for the covariant derivative of $\phi$ containing couplings with the gauge bosons which just reduces to the normal derivative $D_\mu\rightarrow \partial_\mu$ (since during inflation, there are no fields other than the inflaton).

In order to simplify the calculations, we make the following conformal transformation to write the action $S_J$ in the Einstein frame~\cite{Capozziello:1996xg, Kaiser:2010ps}:
\begin{equation}
\hat{g}_{\mu\nu}=\Omega^{2}g_{\mu\nu},~~\sqrt{-\hat{g}}=\Omega^{4}\sqrt{-g},
\end{equation}
so that it looks like a regular field theory action with no explicit couplings to gravity. In the above transformation, $\hat{g}$ represents the metric in the Einstein frame. To make the kinetic term of the inflaton canonical, we redefine $\phi$ by
\begin{equation}
\frac{d\chi}{d\phi}=\sqrt{\frac{\Omega^{2}+\frac{6\xi^{2}\phi^{2}}{M_{P}^{2}}}{\Omega^{4}}}=Z(\phi),
\end{equation}
where $\chi$ is the canonical field. Using these inputs, the inflationary potential in the Einstein frame can be written as, 
\begin{equation}
V_{E}(\phi(\chi))=\frac{V_{J}\big(\phi(\chi)\big)}{\big (\Omega\big(\phi(\chi)\big)\big )^{4}}=\frac{1}{4}\frac{\lambda_2\phi^4}{\big (1+\frac{\xi\phi^2}{M_P^2}\big )^2},
\end{equation}
where $V_J(\phi)$ is identical to $V_{\rm Inf}(\phi)$ in equation (\ref{eq:PotPhi}). We then make another redefinition: $\Phi={\frac{\phi}{\sqrt{1+ \frac{\xi\phi^2}{M_P^2}}}}$ and reach at a much simpler from of $V_E$ given by
\begin{align}
 V_E(\Phi)=\frac{1}{4}\lambda_2\Phi^4.\label{eq:InfE}
\end{align}
Note that for an accurate analysis, one should work with renormalisation group (RG) improved potential and in that case, $\lambda_2$ in equation (\ref{eq:InfE}) will be function of $\Phi$ such that,
\begin{align}
 V_E(\Phi)=\frac{1}{4}\lambda_2(\Phi)\Phi^4.
\end{align}
The one loop renormalisation group evolution (RGE) equations of the relevant parameters associated with the inflationary dynamics are given by,
\begin{align}
\beta_{\lambda_2}&= (18 s^2+2)\lambda_2^2 + 2 \lambda_3^2-\Big(48 g_{BL}^2-2\Sigma_N^2\Big)\lambda_2+96g_{BL}^{4}-\Sigma_N^4,\label{eq:RG1}\\
\beta_{\xi}&=\Big(\xi +\frac{1}{6}\Big)\Big((1+s^2\lambda_2)-2\zeta\Big),\\
\beta_{g_{BL}}&=\Big(\frac{32+4s}{3}\Big)g_{BL}^3,\\
\beta_{Y_{N_i}}&=Y_{N_i}^3-6g_{BL}^2Y_{N_i}+\frac{1}{2}Y_{N_i}\Sigma_N^2,
\end{align}
where we define $s=\left (1+\frac{\xi\phi^2}{M_P^2}\right )\left(1+(1+6\xi)\frac{\xi\phi^2}{M_P^2} \right)^{-1}$, $\zeta=\frac{1}{(4\pi)^2}\Big(\frac{1}{2}\Sigma_N^2-12g_{BL}^2\Big)$, $\Sigma_N^2=\sum_{i=1}^3Y_{N_i}^2$ and $\Sigma_N^4=\sum_{i=1}^3Y_{N_i}^4$ and $\beta_{x_i}=\frac{1}{16\pi^{2}}\frac{dx_i}{d~{\rm ln}\Phi}$. The RGE equations for rest of the couplings are provided in Appendix \ref{appen1}.

\begin{figure}[h]
\includegraphics[scale=0.42]{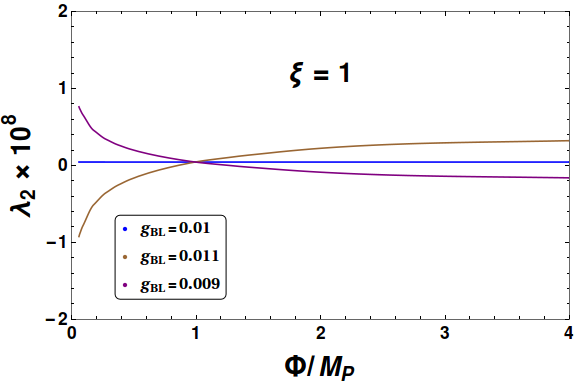}~~\includegraphics[scale=0.42]{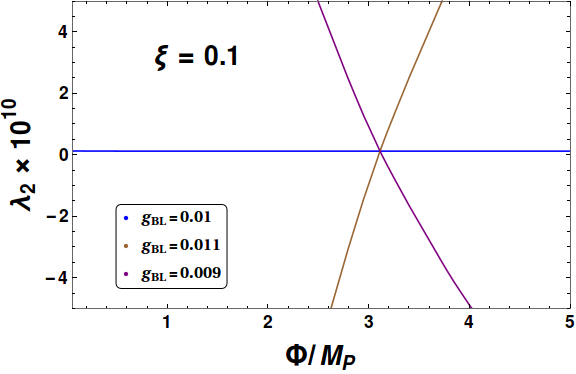}
\caption{RG running of $\lambda_2$ as function of $\Phi$
considering the stability condition (blue) $\Delta\sim0$ with  $\xi=1$ (left panel)
and $\xi=0.1$ (right panel). Brown and purple curves show $\pm 10\%$ variation from $\Delta\sim0$.}
\label{fig:Variation-of-l}
\end{figure}

We choose the heavy neutrino mass spectrum, satisfying the hierarchy $M_{N_1}\ll M_{N_2}<M_{N_3}$ and a diagonal RH neutrino mass matrix. Note that, from this section onwards, we are denoting the RHNs as $N_i$ only without denoting the chirality explicitly. For simplicity, we denote $Y_{N_{22}} \equiv Y_{N_2}, Y_{N_{33}} \equiv Y_{N_3}$. Thus the right handed neutrino mass hierarchy implies $Y_{N_1} \ll Y_{N_2} <Y_{N_3}$. Let us first analyse the case where the RG running
of $\lambda_2$ is dominated by $g_{BL}$ and $Y_{N_{2,3}}$. Then equation (\ref{eq:RG1}) can be rewritten as,
\begin{align}
 \beta_{\lambda_2}\simeq 96g_{BL}^{4}-
 Y_{N_2}^{4}-Y_{N_3}^{4}+2\lambda_3^2.\label{eq:RG2}
\end{align}
We ignore the contributions of $\lambda_{2}$ and $Y_{N_1}$ in the R.H.S. of equation (\ref{eq:RG2}) considering them to be negligible\footnote{Unless the non-minimal coupling $\xi$ is very large, the self-quartic coupling of inflaton must be very small in order to be in agreement with correct inflationary parameters \cite{Okada:2010jf}.}. Since $\lambda_2$
is very small,  $\beta_{\lambda_2}\ll0$ or $ \beta_{\lambda_2}\gg 0$ can cause sharp changes in $\lambda_2$ value from its initial magnitude during the
evolution. It may also happen that $\lambda_2$ becomes negative at some energy scale. Then the inflationary potential would
turn unstable along $\phi$ field direction. Therefore  the most acceptable
case is to make $\beta_{\lambda_2}\rightarrow0$ at least during inflation so that the inflationary potential remains stable \cite{Okada:2015lia}. To ensure $\beta_{\lambda_2}\simeq0$, the equality $\Delta=96g_{BL}^{4}-82Y_{N_2}^4+2\lambda_3^2\sim0$ has to be maintained, where we have assumed $Y_{N_3}=3Y_{N_2}$. We can further simplify the expression for $\Delta$ by assuming $\lambda_3^2\ll g_{BL}^4$. In Fig. \ref{fig:Variation-of-l}, we show the RG running of $\lambda_2$
as a function of $\Phi$ for different values of $g_{BL}$ considering (left panel) $\xi=1$ and (right panel) $\xi=0.1$. The $\lambda_2$ running for $\Delta\sim 0$ is shown in blue colour while the other colours represent the cases where the $\Delta\sim 0$ condition gets violated by $\pm 10\%$. Fig. \ref{fig:Variation-of-l} clearly points out that indeed a small violation of the $\Delta\sim 0$ criteria can cause sharp instability of the inflationary potential.

\begin{figure}[h]
\includegraphics[scale=0.4]{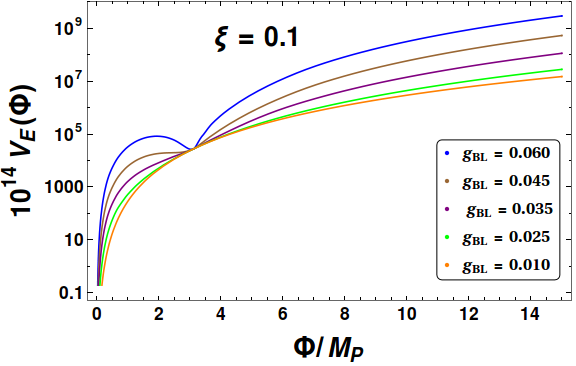} \includegraphics[scale=0.39]{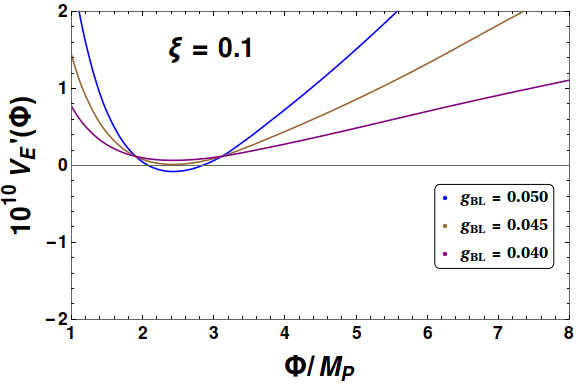}\\
\includegraphics[scale=0.4]{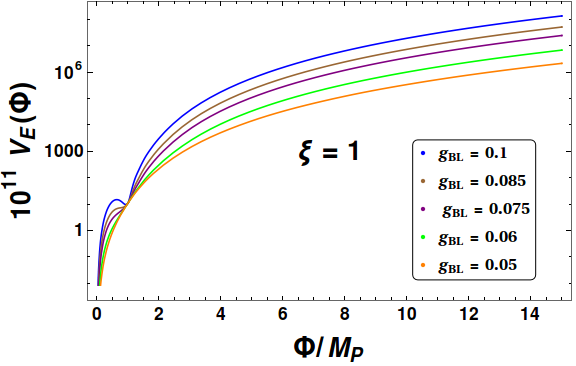} \includegraphics[scale=0.39]{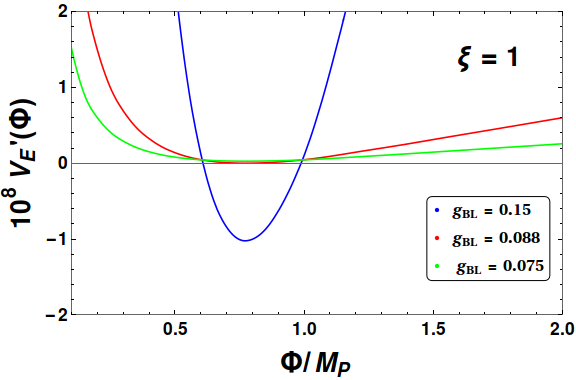}
\caption{ (Left) The inflationary potential and (right) first derivative of the inflationary potential are plotted for
different values of $g_{BL}$ considering $\Delta\sim 0$ with $\xi=0.1$ (top) and $\xi=1$ (bottom).}
\label{fig:The-potential and its der.}
\end{figure}

\begin{figure}[h]
\noindent \begin{centering}
\includegraphics[scale=0.5]{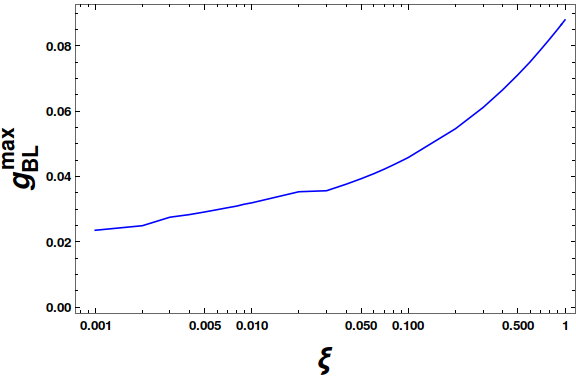}
\par\end{centering}
\caption{\label{fig:gmax vs xi}Variation of $g_{BL}^{\rm max}$ as a function of $\xi$. }
\end{figure}

In upper left panel of Fig. \ref{fig:The-potential and its der.}, we show the behaviour of
the inflationary potential $V_E$ as a function of $\Phi$ for different values
of $g_{BL}$ considering $\xi=0.1$. The value of $\Sigma_N^4$ is determined from
the equality $\Delta$ earlier defined. As it can be observed,
with the increase of $g_{BL}$, the potential starts to develop a local minimum
near some $\Phi$ value say, $\Phi_I$. If such a local minimum exists, then the field could be trapped
there and the inflaton will stop rolling. This  provides an upper bound
on $g_{BL}$ such that the local minimum of $V_E(\Phi)$ does not appear. The existence of a local minimum can be further confirmed
if $\frac{dV_E(\Phi)}{d\Phi}\simeq0$ near $\Phi_{I}$. This condition can
be rewritten as \begin{equation}
\frac{dV_E}{d\Phi}=\frac{\beta_{\lambda_2}}{4}+\lambda_2(\Phi)\simeq0\label{eq:6}
\end{equation}
We plot $\frac{dV_E}{d\Phi} = V^{\prime}_E (\Phi)$ in upper right panel of Fig. \ref{fig:The-potential and its der.}
as a function of $\Phi$. We observe that for $g_{BL}\gtrsim g_{BL}^{\rm max}=0.045$, the inflationary
potential indeed develops a local minimum near $\Phi_I=4 M_P$. Similar conclusion can be drawn for $\xi=1$ as shown in lower panel of Fig. \ref{fig:The-potential and its der.} .
One important point to be noted is that the value of $g_{BL}^{\rm max}$ gets enhanced with the increase of $\xi$. We illustrate this in Fig. \ref{fig:gmax vs xi} where $g_{BL}^{\rm max}$ is plotted against different values of $\xi$.

\begin{figure}[h]
\noindent \begin{centering}
\includegraphics[scale=0.37]{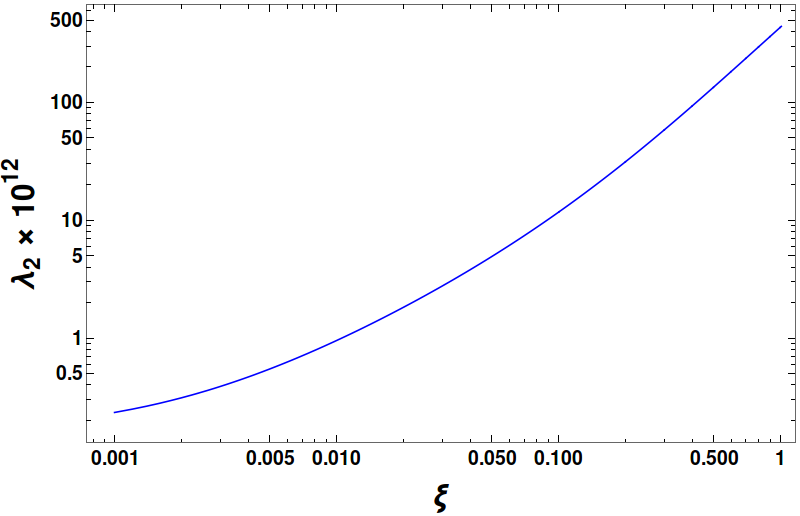}
\par\end{centering}
\caption{\label{fig:xi vs l}Variation of $\lambda_2$ as a function $\xi$ in order to produce the correct amount of curvature perturbation spectrum $P_s$. }
\end{figure}

\begin{figure}[h]
\includegraphics[scale=0.38]{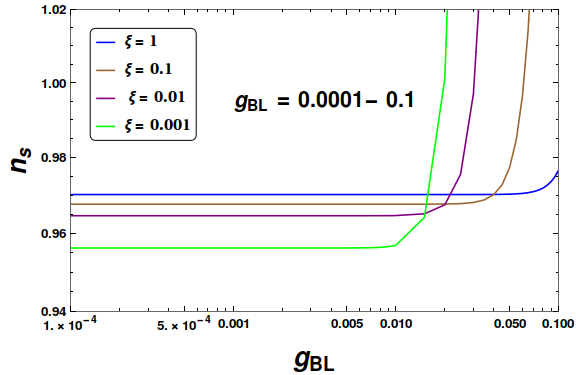}$\,\;$ \includegraphics[scale=0.38]{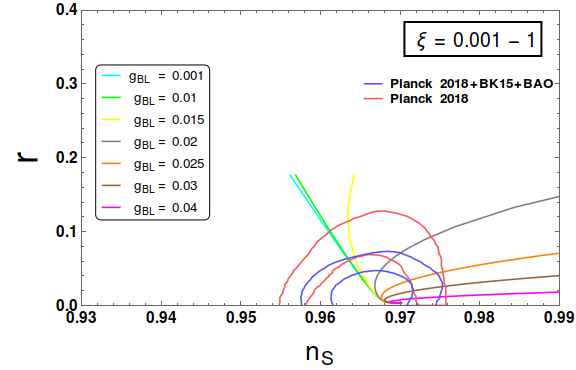}
\caption{\label{fig:gvsns} [Left] The magnitude of spectral index $n_{s}$ is plotted against $g_{BL}$
for different $\xi$s. [Right] $n_{s}-r$ contours for different set of constant $g_{BL}$ values with $\xi=0.001-1$. The $1\sigma$ and $2\sigma$ bounds from Planck 2018+BAO+BK15 are also included.}
\end{figure}

\begin{figure}[h]
\includegraphics[scale=0.27]{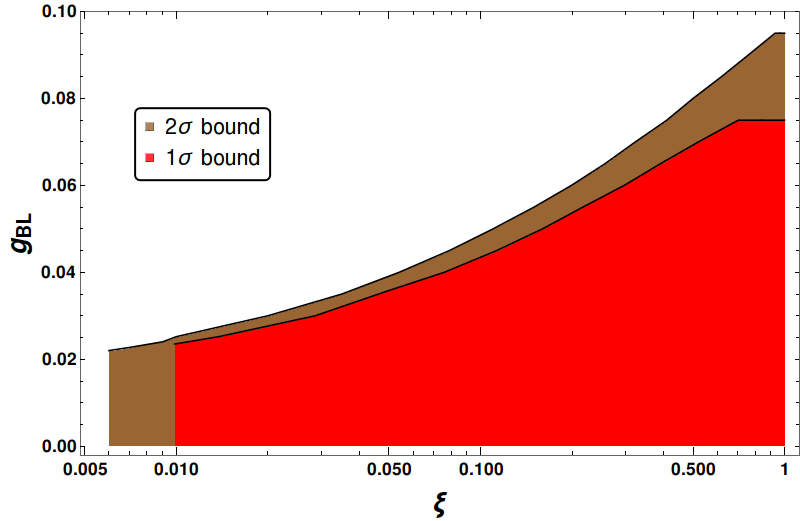}~
\includegraphics[scale=0.39]{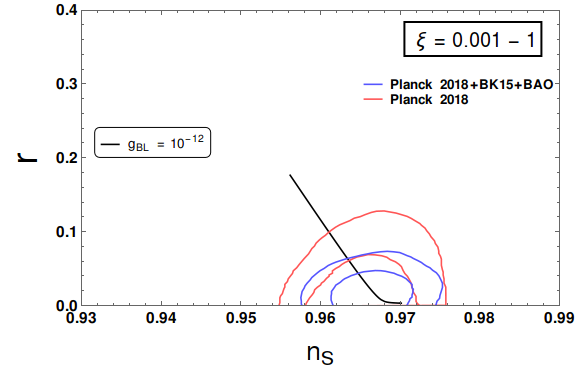}
\caption{\label{fig:-nsr}[Left]: Allowed parameter space from inflation in $g_{BL}-\xi$ plane by Planck 2018 $1\sigma$ and $2\sigma$ bounds. [Right] $n_s-r$ contour by varying $\xi$ and considering $g_{BL}^4,\Sigma_N^4\ll\lambda_2^2$ at inflationary energy scale for $N_e=60$.}
\end{figure}

Next, we move on to calculate the predictions for inflationary observables. In terms of the original field $\phi$, the slow roll parameters ($\epsilon, ~\eta$) and number of e-folds ($N_e$) are found to be 
\begin{align}
&\epsilon(\phi)=\frac{M_{P}^{2}}{2Z(\phi)^{2}}\Bigg(\frac{V_{E}'(\phi)}{V_{E}(\phi)}\Bigg)^{2},\\
&\eta(\phi)=\frac{M_{P}^{2}}{Z(\phi)^{2}}\Bigg(\frac{V_{E}''(\phi)}{V_{E}(\phi)}-\frac{V_{E}'(\phi)Z'(\phi)}{V_{E}(\phi)Z(\phi)}\bigg), \\
&N_e=\int_{\phi_{t}}^{\phi_{end}}\frac{Z^{2}V_{E}(\phi)}{V_{E}'(\phi)}\frac{{d}\phi}{M_{\mathrm{P}}},
\end{align}
respectively. The inflationary observables such as spectral index ($n_s$), tensor to scalar ratio ($r$) and scalar perturbation spectrum ($P_S$) can be expressed in terms of the slow roll parameters as
\begin{align}
 n_s=1-6\epsilon+2\eta,~~r=16\epsilon,~~P_S=\frac{V_E(\phi)}{24M_P^4\pi^2\epsilon}.
\end{align}
All these quantities have to be determined at the horizon exit of the inflaton ($\phi_t$) and we consider the number of e-folds $N_e=60$ for the numerical analysis. We perform a numerical scan over $g_{BL}$ and $\xi$ to estimate the inflationary observables $n_s$ and $r$ considering $\Delta\sim 0$. The initial value of $\lambda_2$ is determined to
produce the correct observed value of scalar perturbation spectrum
$P_{S}$ at horizon exit.  In Fig. \ref{fig:xi vs l} we show the variation of $\lambda_2$ with $\xi$
to be consistent with the observed value of $P_{S}=2.4\times10^{-9}$.
It turns out that the value of $r$ does not change much with the variation of
$g_{BL}$ for a constant value of $\xi$ since $\beta_\lambda=0$ at inflationary energy scale. Contrary to this, value of $n_s$
is quite sensitive to $g_{BL}$. We see from left panel of Fig. \ref{fig:gvsns} that  $n_s$ increases with the enhancement of $g_{BL}$ for different values of $\xi$. The rate of increase of $n_s$ with $g_{BL}$ turns flatter with the rise of $\xi$ value.
In the right panel of Fig. \ref{fig:gvsns} we plot $n_s-r$ contours for different $g_{BL}$ values and by  varying $\xi$ in the range 0.001-1. For comparison purpose we also insert the Planck 2018+BAO+BK15 1$\sigma$ and $2\sigma$ bounds \cite{Akrami:2018odb}. It is evident that the present setup is able to provide set of $n_s-r$ values, consistent with the experimental constraints.
Finally, in the left panel Fig. \ref{fig:-nsr}, we constrain the $\xi-g_{BL}$ plane which correctly produces the $n_s-r$ values consistent with Planck $1\sigma$ (red) and $2\sigma$ (brown) bounds.

So far we have discussed the case where $g_{BL}^4,\Sigma_N^4\gg \lambda_2^2$ at inflationary energy scale. Hence, it is obvious to consider the opposite limit of these parameters. When $g_{BL}^4,\Sigma_N^4\ll \lambda_2^2$, automatically the inflation scenario merges with the case of quartic inflation and non minimal coupling of inflaton to gravity as originally studied in \cite{Okada:2010jf}. For completeness purpose we discuss this particular case in right panel of Fig. \ref{fig:-nsr} in $n_s-r$ plane. As it is seen the $n_s-r$ contour can still satisfy the Planck 2018 $1\sigma$ bounds for $N_e=60$. The contour of observed value of $P_S$ in $\xi-\lambda_2$ plane remains same as in Fig. \ref{fig:xi vs l}.

\subsection{Reheating}\label{sec:reheat}
Once inflation ends, the thermalisation of the universe, leading to a radiation dominated universe has to be ensured. This is the reheating epoch~\cite{Allahverdi:2010xz}, which takes the universe from the inflationary phase to the radiation-dominated phase. 

Originally, the reheating process was proposed as the perturbative decay of inflaton field into lighter degrees of freedoms  \cite{Albrecht:1982}. During oscillation, the energy of inflaton gets transferred into the relativistic lighter decay products. Approximately, the amount of energy density of the radiation bath is obtained as $\sim 3 M_P^2 \Gamma_\Phi$ where $\Gamma_\Phi$ is the total decay width of inflaton. Considering inflaton decay into radiation only while setting up thermodynamic equilibrium quickly after the decay, the maximum reheating temperature of the universe is found to be 
\begin{align}
T_R\sim \left(\frac{90}{g_*\pi^2}\right)^{1/4}\sqrt{\Gamma_\Phi M_P},
\end{align}
where $g_*$ is the number of relativistic degrees of freedom in the thermal bath.

However, the success of this perturbative decay mechanism of inflaton is somewhat limited. In initial stages of reheating, the phenomena of parametric resonance might be important and may lead to explosive particle production which the theory of perturbative reheating does not take into account. This dynamics is known as preheating  \cite{Kofman:1994, Kofman:1997, Greene:1997}. In particular, if the oscillation amplitude of the inflaton is sufficiently large, the number density of the produced bosonic particles might be enhanced ($n_{k}\gg 1$) due to the effects related to Bose statistics. In an expanding universe, this process occurs in a stochastic manner, and is known as stochastic resonance. The produced particles, due to the large amplitude of inflaton, turn non-relativistic and further decay into lighter relativistic particles. The primary condition which needs to be satisfied to attain parametric resonance in an inflationary framework is that the decay width of non-relativistic particles should be less than its production rate. Parametric resonance halts once the inflation oscillation amplitude becomes small and the resonance becomes narrower.

The presence of parametric resonance as described above could raise the final reheating temperature compared to the one obtained by considering the perturbative reheating only. However, if the couplings of the inflaton with the lighter particles are not strong enough, the resonance is narrow or not broad enough. This makes preheating inefficient. In particular, it was shown in ref. \cite{Kofman:1997, Greene:1997} that for couplings $\lesssim \mathcal{O}(10^{-4})$ the broad resonance does not take place (resulting $n_k\ll 1$ \cite{Kofman:1997, Felder:1999}) and preheating finishes at very early stage without posing significant impact on the final reheating temperature. In that case the reheating temperature of the universe is dominantly guided by the perturbative reheating. 
 
From the inflationary perspective, we are having two different kind of  scenarios  having phenomenological relevance namely, (i) $g_{BL}^4, \Sigma_N^4\gg\lambda_2^2$ and (ii) $g_{BL}^4, \Sigma_N^4\ll\lambda_2^2$.

\begin{figure}[h]
\includegraphics[height=5.5cm,width=7.9cm]{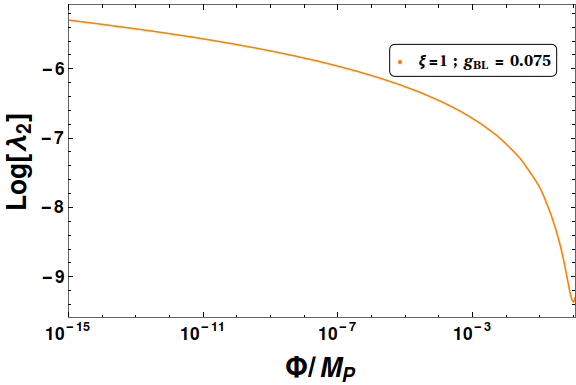}~
\includegraphics[height=5.5cm,width=7.9cm]{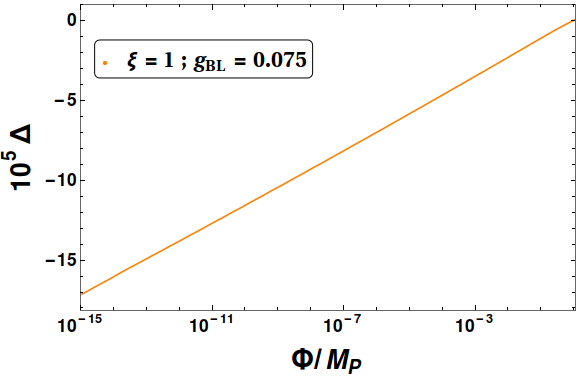}
\caption{RG running of $\lambda_2$ (left) and $\Delta$ (right) as function of the energy scale $\Phi$ considering $\xi=1$ and $g_{BL}=0.075$.}
\label{fig:RGPlot}
\end{figure}

For the first case $g_{BL}$ is large and thus $\Delta\sim 0$ is an essential condition for the stability of inflationary potential. We consider $\lambda_3\ll g_{BL}^2$ so that it does not effect the evolution of $\Delta$ significantly. This assumption was made earlier also while determining the fate of inflation. The value of $\Delta$ as defined earlier changes by small amount in its RG evolution (see right panel of Fig. \ref{fig:RGPlot}). It is found that the value of $\lambda_2(\Phi_I)$ changes by order of magnitudes at low scale, for example $\lambda_2(\Phi=1 {\rm~ TeV})$ becomes $\mathcal{O}(10^{-6})$ from $4.34\times 10^{-10}$ at inflationary scale (considering $\xi=1$, see left panel of Fig. \ref{fig:RGPlot}). During preheating stage, first $Z_{BL}$, SM bosons get produced during the oscillation regime. Afterwards due to inflaton induced large mass these produced $Z_{BL}$ and SM bosons turn non-relativistic, and they decay into the lighter relativistic particles. In a whole, this particular process comprises of unusual stochastic resonance production of lighter non relativistic particles, their further decays, backreaction in the presence of an expanding universe. Hence the estimate of the correct reheating temperature is more involved and requires rigorous lattice simulation \cite{Garcia:2009, Maity:2019}. Since we shall see in a while that this scenario turns out to be disfavoured due to overproduction of WIMP DM relic, we do not elaborate on this further\footnote{In refs. \cite{Bezrukov:2009, Borah:2018rca} a detailed analysis on preheating in a similar setup has been performed considering $\xi\gg 1$.}. 

\begin{figure}[h]
\includegraphics[height=10cm,width=12cm]{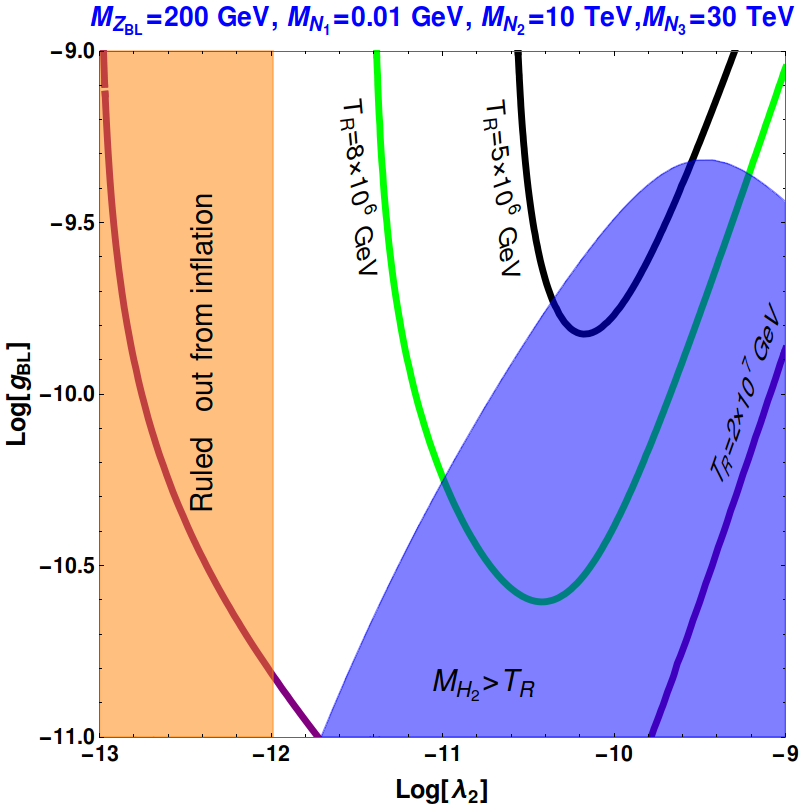}~`
\caption{Case II: Contours of $T_R$ in $g_{BL}-\lambda_2$ plane considering fixed values of $M_{Z_{BL}}, M_{N_{1,2,3}}$. The orange region is ruled out from inflation and in the blue region mass of the inflaton is larger than the reheating temperature.}
\label{fig:reheatII}
\end{figure}

In the second case $g_{BL}^4, \Sigma_N^4\ll\lambda_2^2$, the inflationary potential is mainly driven by $\lambda_2$ with other couplings sufficiently small. Hence $\Delta\sim 0$ is not a necessary condition for this case. However the coupling $\lambda_3$ (we take $\mathcal{O}(10^{-10})$) should be still much smaller than unity so that the stability of inflation potential remains intact. Here, due to the smallness of all relevant couplings there will not be any significant changes during their RG running unlike in the earlier case. The important point is with the estimates of $g_{BL}$ and $\lambda_3$ from inflation, the preheating stage never turns efficient and gets over at very early stage of inflaton oscillation. Then the reheating of the universe will be effectively dictated by the perturbative decay of inflaton. Here, depending on the mass scale (or $\lambda_2$), the tree level decay of inflaton into $Z_{BL}Z_{BL},~H_1H_1$ final states are possible. The inflaton can also decay into right handed neutrinos, if kinematically allowed. In Fig. \ref{fig:reheatII} we show the contours of different values of $T_R$ (ranging from $5\times 10^6~{\rm GeV}-2\times 10^7$ GeV) in $g_{BL}-\lambda_2$ plane. For this purpose we fix $M_{Z_{BL}}=200$ GeV, $M_{N_1}=10 {~\rm MeV},~M_{N_2}=10$ TeV and $M_{N_3}=30$ TeV. The orange coloured region is ruled out from the requirement of reproducing the observed value of scalar perturbation spectrum
$P_{S}$ at horizon exit. In the blue coloured region inflaton mass turns larger than the reheating temperature and hence it remains out of equilibrium. This may have important implications for other related phenomenology as we will discuss in a while. 
\section{Dark matter}\label{sec:DM}
In this section, we discuss the dark matter phenomenology in detail and attempt to find its consistency with the inflationary dynamics. As mentioned earlier, $N_1$ is the DM candidate which is odd under $Z_2$ and hence stable. For earlier studies of DM in this model, one may refer to \cite{Okada:2010wd, Basak:2013cga, Okada:2016gsh, Okada:2018ktp, Escudero:2018fwn}. While the $Z_2$ odd RHN is the DM candidate, the other two RHN's take part in the usual type I seesaw mechanism, giving rise to light neutrino masses and mixing. Since DM is a singlet under SM gauge symmetry, it can interact with the visible sector particles only via gauge $(Z_{BL})$ or scalar $(H_{1,2})$ interactions. Now, depending upon the two cases namely, (i) $g_{BL}^4, \Sigma_N^4\gg\lambda_2^2$ and (ii) $g_{BL}^4, \Sigma_N^4\ll\lambda_2^2$ discussed in the context of inflation, DM-SM couplings can either be of order unity or very small. This will lead to completely different DM phenomenology namely, thermal or WIMP type and non-thermal or FIMP type, which we discuss separately below.

\begin{figure}[h]
\includegraphics[height=3.3cm,width=5cm]{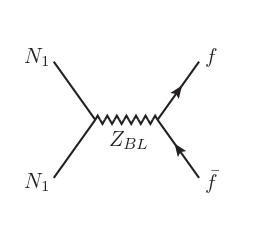}
\includegraphics[height=3.3cm,width=5cm]{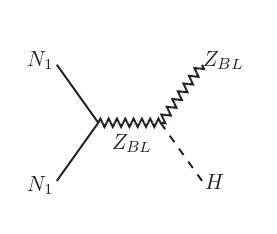}
\includegraphics[height=3.3cm,width=5cm]{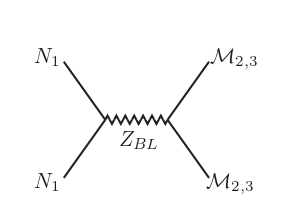}
\includegraphics[height=3.3cm,width=5cm]{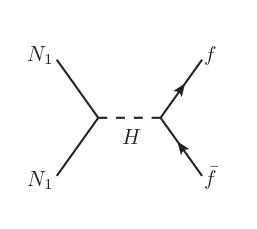}
\includegraphics[height=3.3cm,width=5cm]{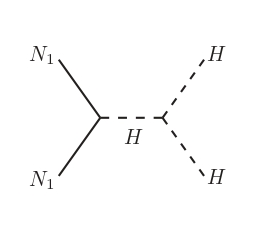}
\includegraphics[height=3.3cm,width=5cm]{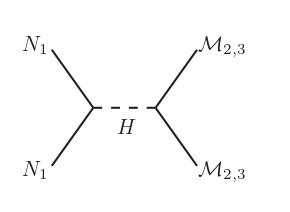}
\includegraphics[height=3.3cm,width=5cm]{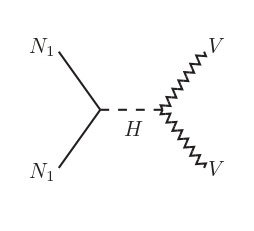}
\includegraphics[height=3.3cm,width=5cm]{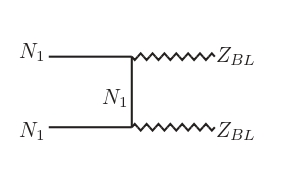}
\includegraphics[height=3.3cm,width=5cm]{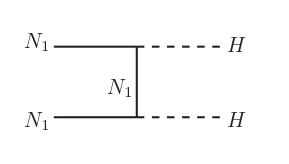}
\caption{All possible annihilation processes of DM ($N_1$) into various final state particles. Here, $\mathcal{M}_{2,3}$, $H$ and $V$ represent the Majorana neutrinos ( $N_{2,3}$ or $\nu_{2,3}$), scalars $H_{1}$, $H_{2}$ and electroweak vector bosons respectively.}
\label{fig:FeynD}
\end{figure}

For the first case, that is, $g_{BL}^4,\Sigma_N^4\gg\lambda_2^2$, it is expected that the DM stays in thermal equilibrium with the SM particles in the early universe and thus falls into the WIMP category. The DM can annihilate into different final states in the thermal bath through processes mediated by scalars and the $U(1)_{B-L}$ gauge boson. In Fig. \ref{fig:FeynD}, we exhibit the possible annihilation processes of $N_1$ in the present framework. Please note that, in principle, the symmetry of the model allows a kinetic mixing term between $U(1)_Y$ of SM and $U(1)_{B-L}$ of the form $\frac{\varepsilon}{2} B^{\alpha \beta} B^{\prime}_{\alpha \beta}$ where $B^{\alpha\beta}=  \partial^{\alpha}B^{\beta}-\partial^{\beta}B^{\alpha}$ and $\varepsilon$ is the mixing parameter. Even if we turn off such mixing at tree level as we have done here, one can generate such mixing at one loop level since there are particles in the model which are charged under both $U(1)_Y$ and $U(1)_{B-L}$. Such one loop mixing can be approximated as $\varepsilon \approx g_{\rm BL} g_2/(16 \pi^2)$ \cite{Mambrini:2011dw}. Since $g_{\rm BL}$ has tight upper bound from inflationary dynamics, the one loop mixing can be neglected in comparison to other relevant couplings and processes. Therefore, for simplicity, we ignore such kinetic mixing for the rest of our analysis.

\subsection{WIMP DM Scenario}

The evolution of comoving number density of DM ($Y_{\rm DM}=n_{\rm DM}/s$) is determined by the corresponding Boltzmann equation
\begin{align}
 \frac{dY_{\rm DM}}{dz}=-\frac{z\langle\sigma v\rangle s}{\mathcal{H}(M_{N_1})}(Y_{\rm DM}^2-Y_{\rm DM}^{\rm eq^2}),\label{eq:BoltzDM1}
\end{align}
where 
\begin{align}
 Y_{\rm DM}^{\rm eq ^2}=\frac{45}{4\pi^4}\frac{g}{g_{*s}}z^2K_2(z),
\end{align}
with $g$ and $g_{*s}$ being the internal degrees of freedom of the dark matter and relativistic entropy degrees of freedom respectively and $z=M_{N_1}/T$. The $\langle\sigma v\rangle$ in equation (\ref{eq:BoltzDM1}) stands for the thermally averaged cross section of DM annihilation, given by~\cite{Gondolo:1990dk}
\begin{equation}
\langle \sigma v \rangle \ = \ \frac{1}{8M_{N_1}^4T K^2_2\left(\frac{M_{N_1}}{T}\right)} \int\limits^{\infty}_{4M_{N_1}^2}\sigma (s-4M_{N_1}^2)\sqrt{s}\: K_1\left(\frac{\sqrt{s}}{T}\right) ds \, ,
\label{eq:sigmav}
\end{equation}
where $K_i(z)$'s are modified Bessel functions of order $i$.  $\mathcal{H}(M_{N_1})$ represents the Hubble parameter at $T=M_{N_1}$. 

\begin{figure}[h]
\noindent \begin{centering}
\includegraphics[scale=0.35]{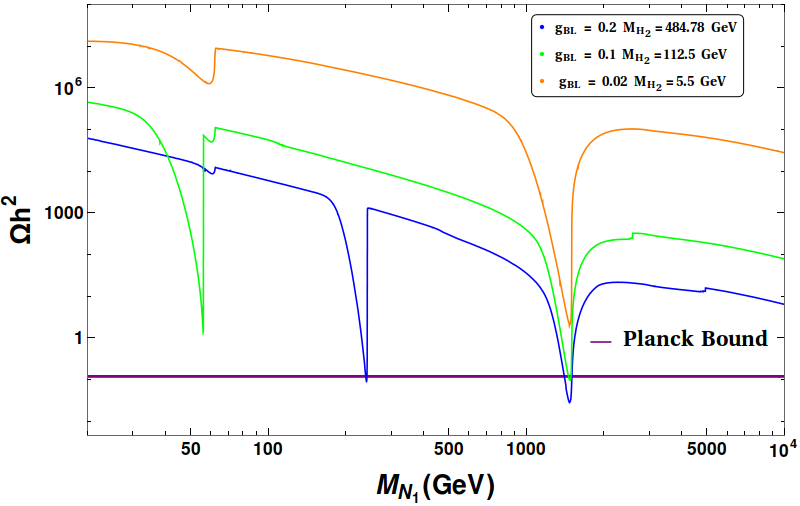}
\par\end{centering}
\caption{\label{fig:Relic plot2}DM Relic as a function of its mass for different set of $g_{BL}$ values with $M_{Z_{BL}}=3$ TeV. We have considered $\lambda_2=4.35\times 10^{-10}$ and $\lambda_3\sim10^{-6}$ at inflationary energy scale.}
\end{figure}
\par

\begin{figure}[h]
\noindent \begin{centering}
\includegraphics[scale=0.35]{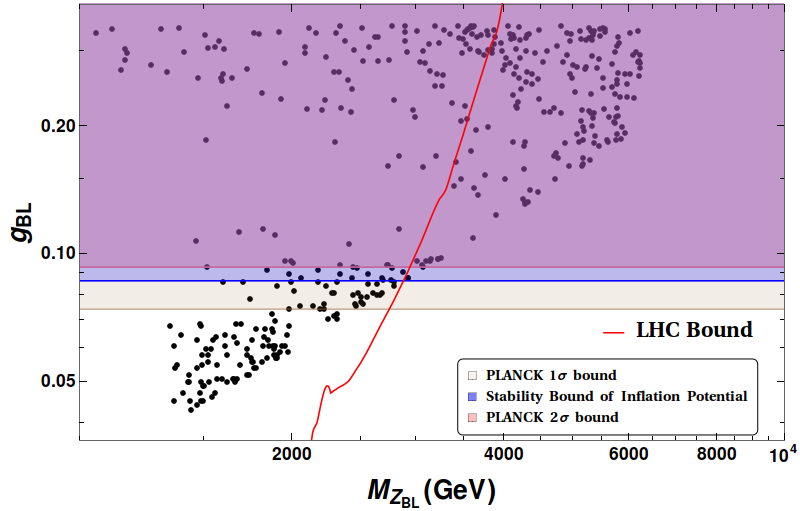}
\par\end{centering}
\caption{\label{fig:Parameter-space}Parameter space satisfying DM relic abundance in $g_{BL}-M_{Z_{BL}}$ plane by considering $\lambda_2=4.35\times 10^{-10}$ and $\lambda_3\sim10^{-6}$ at inflationary energy scale. Bounds arising from LHC, Planck constraints on inflation (1$\sigma$ and 2$\sigma$) and stability of inflationary potential are also shown. The shaded regions are
disallowed. }
\end{figure}

We implement the model in \texttt{FeynRules} \cite{Alloul:2013bka} and then use \texttt{micrOMEGAs} package \cite{Belanger:2013oya} to estimate the relic abundance of DM numerically. The independent parameters which participate in determining the DM relic abundance are the following:
\begin{align}
 \Big\{Y_{N_{2,3}}, Y_{D_{ij}},M_{Z_{BL}},g_{BL},M_{H_2},M_{N_1},\sin\theta\Big\}.
\end{align}
In our case, we have considered the $Y_N$ matrix diagonal and the sum of fourth power of each diagonal elements are fixed by inflationary requirements. However, for the DM analysis we need the magnitude of each individual elements. For simplification purpose we make the choice  $Y_{N_3}=3 Y_{N_2}$ at the inflationary energy scale, to reduce the number of free parameters. The $\Delta\sim 0$ condition was essential at the inflationary energy scale and hence for the DM analysis we need to run the RGE equations of $g_{BL}$ and $Y_{N}$ along with $\lambda_2$ and $\lambda_3$, with the initial condition $\Delta=0$, to estimate their values around few TeV scale, relevant for DM freeze-out. The value of $Y_{N_1}$ will be fixed from the choice of DM mass and then the magnitude of $M_{N_{2,3}}$'s can be computed using $Y_{N_{2,3}}$ values obtained at TeV scale through RG running. Since $Y_{N_1}$ is taken to be  smaller than $Y_{N_{2,3}}$, DM mass $M_{N_{1}}$  is smaller than $M_{N_{2,3}}$'s. We have already discussed the Dirac neutrino Yukawa or $Y_{D}$ matrix and here we use the same form as defined in equation (\ref{casas}) using Casas-Ibarra parametrisation. Here we work with $\lambda_2=4.35\times10^{-10}$ (corresponding to $\xi=1$, see Fig. \ref{fig:xi vs l}), $\lambda_3=10^{-6}$ at inflationary energy scale. 
Since in our working range of gauge coupling $0.01<g_{BL}<0.075$, the reheating temperature $T_R$ is expected to be large, hence it is obvious that the relevant SM and BSM fields will maintain thermal equilibrium with each other.

In Fig. \ref{fig:Relic plot2}, we show the variation of relic as function of DM mass for different set of $g_{BL}$ values (at inflationary energy scale) by keeping $M_{Z_{BL}}$ fixed at 3 TeV. The order of magnitude of $\lambda_2$ and $\lambda_3$ are determined at TeV scale through their RG running corresponding to different $H_2$ mass and $H_2-H_1$ mixing.  With the choices of different mass scales, three resonances appear for $\Omega$ lines at $\frac{M_{H_1}}{2},~\frac{M_{H_2}}{2}$ and 
$\frac{M_{Z_{BL}}}{2}$ respectively. In some cases, one of the scalar resonances is not so prominent due to smallness of $H_2$ mass or $H_2-H_1$ mixing. The purple solid line in Fig. \ref{fig:Relic plot2} represents the observed relic abundance, as per Planck 2018 data \cite{Aghanim:2018eyx}. It is seen that the annihilation through gauge boson is the most efficient one and can satisfy correct relic in two out of three scenarios discussed. 

\begin{figure}[h]
\noindent \begin{centering}
\includegraphics[scale=0.35]{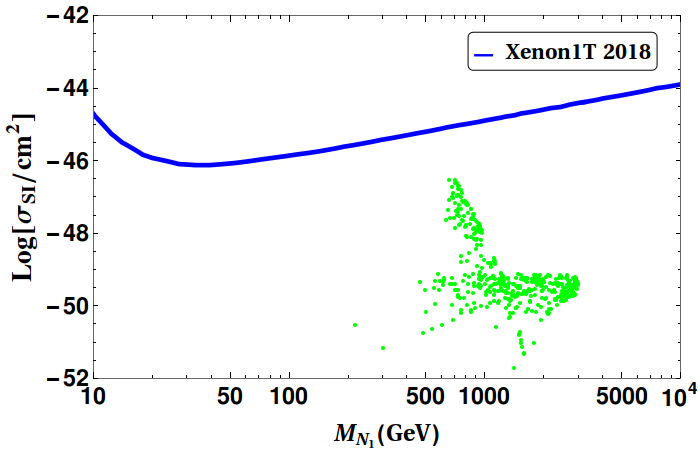}
\par\end{centering}
\caption{\label{fig:Parameter-spaceDD}Direct detection cross sections of the relic satisfied points (green dots) in Fig. \ref{fig:Parameter-space} as function of DM mass is shown along with the bound from XENON1T \cite{Aprile:2017iyp, Aprile:2018dbl}. }
\end{figure}

We then perform a numerical scan to find the parameter space satisfying correct DM relic. In Fig. \ref{fig:Parameter-space}, we display the points satisfying correct DM relic (black dots) in $M_{Z_{BL}}-g_{BL}$ plane considering $M_{Z_{BL}}\lesssim 10$ TeV. We use the values of relevant parameters as earlier mentioned. We also include the LHC bound from dilepton resonance searches  \cite{Aaboud:2017buh} (red curve), Planck constraints on inflation and stability bounds of the inflationary potential for comparison purpose. The shaded regions are disfavoured from the respective constraints. To conclude, we observe that with TeV scale or lower $Z_{BL}$ mass, it is not possible to generate the correct value of relic abundance for WIMP dark matter while being in agreement with LHC and inflationary observables simultaneously. We also check that direct detection limits on spin-independent DM-nucleon cross section from the XENON1T experiment \cite{Aprile:2017iyp, Aprile:2018dbl} and find that such bounds do not put any additional constraint on this parameter space as all the points shown in Fig. \ref{fig:Parameter-spaceDD} obey these bounds.

\subsection{FIMP DM Scenario}
In the second case ($g_{BL}^4,\Sigma_N^4\ll\lambda_2^2$), the couplings responsible for DM-SM interactions are tiny and hence it is expected that DM may never reach thermal equilibrium with the standard bath. This falls under the ballpark of FIMP dark matter, discussed earlier. For earlier work on fermion singlet as FIMP DM in $U(1)_{B-L}$ model, please see \cite{Biswas:2016bfo, Biswas:2016iyh} and references therein. A recent study also discussed the possibility of scalar singlet responsible for breaking $B-L$ gauge symmetry spontaneously to be a long-lived FIMP DM candidate \cite{Mohapatra:2020bze}. If $N_1$ is a FIMP candidate, it can be produced non-thermally, due to decay or annihilation of other particles. In case $Z_2$ symmetry is exact, $N_1$ will be only pair produced as it is the only $Z_2$ odd particle.  All scattering processes shown in Fig. \ref{fig:FeynD} while discussing WIMP scenario can potentially contribute to the production of FIMP DM as well, when considered in the reverse direction. In addition, decays of $H_{1,2}$ and $Z_{BL}$, if kinematically allowed, can also contribute to the relic density of $N_1$. Typically, if same dimensionless couplings govern the strength of both decay and annihilation processes, the former dominates simply due to power counting. This is precisely the scenario here and FIMP is primarily produced from decays.

\begin{figure}[h]
\includegraphics[height=3cm,width=8cm]{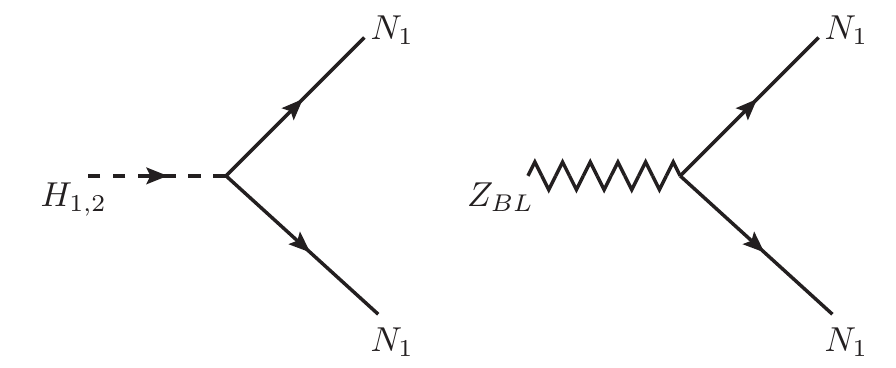}
\caption{DM production channels from tree level decay of heavier particles.}
\label{fig:FeynD2}
\end{figure}

For our numerical calculation, we choose $\lambda_2\sim 1.04\times 10^{-12}$ at inflationary energy scale corresponding to $\xi\sim 0.01$ from inflationary requirements (see Fig. \ref{fig:xi vs l}). Then from Fig. \ref{fig:reheatII}, it is evident that for this choice of $\lambda_2$, $H_2$ would be in thermal equilibrium with other SM particles by virtue of its coupling with Higgs as well as heavy right handed neutrinos $N_{2,3}$ which also maintain equilibrium since their masses considered here are below $T_R$ and they can interact to SM fields through Yukawa interaction. We would like to keep $\lambda_3\sim 10^{-10}$ extremely small so that it does not alter the RG running of $\lambda_2$ during inflation. Since $g_{BL}$ is also very small to justify FIMP nature of DM, we will investigate the possibility of production of $N_1$ DM from non thermal tree level decays of $Z_{BL}$ and $H_2$ (see Fig. \ref{fig:FeynD2}). We will consider two benchmark choices of $M_{Z_{BL}}<10$ TeV for the analysis. It is to be noted that $Z_{BL}$ which interacts only via gauge coupling $g_{BL}$ is also expected to be out of equilibrium. Hence non-thermal production of $Z_{BL}$ from other bath particles and its subsequent decay into $N_1$ pairs play non-trivial roles. We therefore use coupled Boltzmann equations for both $Z_{BL}$ and $N_1$ to calculate the relic abundance of $N_1$ in this scenario.

The evolution of the comoving number densities for $Z_{BL}$ and DM are governed by the following coupled Boltzmann equations \cite{Biswas:2016bfo}
\begin{align}
 \frac{dY_{Z_{BL}}}{dz}=&\frac{2 M_P}{1.66 M_{H_1}^2} \frac{z\sqrt{g_*(z)}}{g_{*s}(z)}\Bigg(\langle\Gamma_{H_{1,2}\rightarrow Z_{BL}Z_{BL}}\rangle(Y_{H_{1,2}}^{\rm eq}-Y_{Z_{BL}})-\langle\Gamma_{Z_{BL}\rightarrow {\rm all} }\rangle Y_{Z_{BL}}\Bigg),\label{FimpB1}\\
 \frac{dY_{\rm DM}}{dz}=&\frac{2 M_P}{1.66 M_{H_{1}}^2} \frac{z\sqrt{g_*(z)}}{g_{*s}(z)}\Bigg(\langle\Gamma_{H_{1,2}\rightarrow N_1 N_1}\rangle(Y_{H_{1,2}}^{\rm eq}-Y_{\rm DM})+\langle\Gamma_{Z_{BL}\rightarrow N_1 N_1 }\rangle (Y_{Z_{BL}}-Y_{\rm DM})\Bigg)\nonumber\\&+\frac{4\pi^2}{45\times 1.66}\frac{g_{*s}}{\sqrt{g_*}}\frac{M_{H_1}M_P}{z^2}
 \times\Bigg\{\langle\sigma v_{xx\rightarrow N_1 N_1}\rangle(Y_{x}^{{\rm eq}^2}-Y_{\rm DM}^2)+\langle\sigma v_{Z_{BL}Z_{BL}\rightarrow N_1 N_1}\rangle(Y_{Z_{BL}}^2-Y_{\rm DM}^2)\Bigg\},\label{FimpB2}
\end{align}
where $z=M_{H_1}/T$ and $x$ represents all possible initial states. $g_{*}(z)$ is defined by
\begin{eqnarray}
\sqrt{g_{\star}(z)} = \dfrac{g_{*\rm s}(z)}
{\sqrt{g_{\rho}(z)}}\,\left(1 -\dfrac{1}{3}
\dfrac{{\rm d}\,{\rm ln}\,g_{*\rm s}(z)}{{\rm d} \,{\rm ln} z}\right)\,
\end{eqnarray}
while $g_{*s}$ is same as defined earlier. Here, $g_{\rho}(x)$ denotes the effective number of degrees
of freedom related to the energy density of the universe at $z$. The $\langle\Gamma_{A\rightarrow BC}\rangle$ denotes the thermally averaged decay width which is given by
\begin{align}
 \langle\Gamma_{A\rightarrow BC}\rangle=\frac{K_1(z)}{K_2(z)}\Gamma_{A\rightarrow BC}.
\end{align} 
\begin{figure}[h]
\noindent \begin{centering}
\includegraphics[height=7cm,width=8cm]{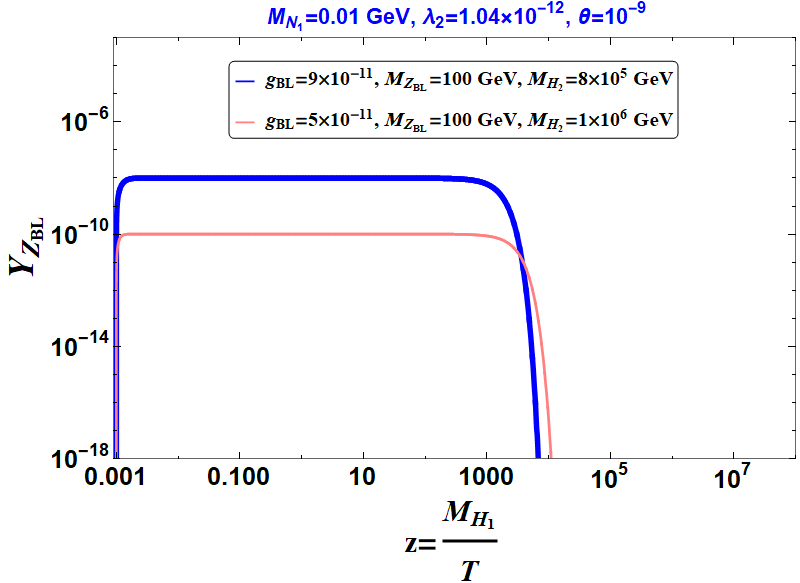}
\includegraphics[height=7cm,width=8cm]{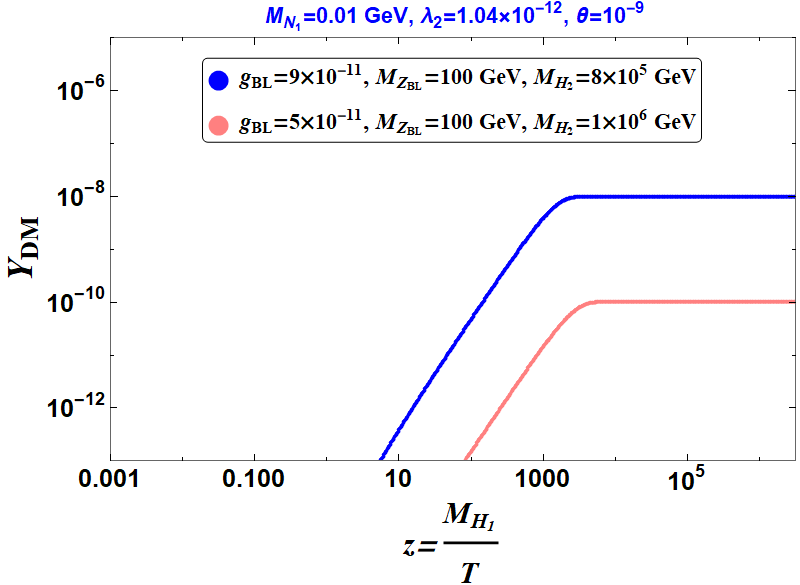}
\par\end{centering}
\caption{Evolution of comoving number densities of $Z_{BL}$ (left panel) and DM $N_1$ (right panel) as function of temperature.}
\label{fig:FimpY}
\end{figure}
\noindent Since initial densities of both $Z_{BL}$ and $N_1$ are almost vanishing, one can ignore $Y_{Z_{BL}}$ and $Y_{\rm DM}$ from first term within each bracket on right hand side of equations (\ref{FimpB1}) and (\ref{FimpB2}). 

In left panel of Fig. \ref{fig:FimpY}, we show the evolution of $Y_{Z_{BL}}$ against $z$ for benchmark choices of $g_{BL}$ and other relevant parameters indicated in the figure. It is seen that $Y_{Z_{BL}}$ starts from a vanishingly small value initially and reaches a sizeable value with the lowering of temperature very quickly. The initial increase in $Z_{BL}$ abundance happens primarily from $H_2$ decays. As expected, the production of $Z_{BL}$ from $H_2$ decay becomes efficient around $T \sim M_{H_2}$ which corresponds to $z = M_{H_1}/T \sim 10^{-3}$. For $T< M_{H_2}$ there is a Boltzmann suppression in the equilibrium abundance of $H_2$ which makes $Z_{BL}$ production less efficient leading to the plateau region where $Y_{Z_{BL}}$ remains more or less constant. We also observe that a larger value of $g_{BL}$ while keeping $M_{Z_{BL}}$ fixed gives  larger yield for $Z_{BL}$. The reason behind this is two-fold. Firstly, the partial decay width of $H_2$ into $Z_{BL}$ pairs rises with the increase in $g_{BL}$ values for our chosen benchmark points. Note that this partial decay width is function of $g_{BL}, M_{H_2}$ and can be expressed as (in the limit $M^2_{Z_{BL}} \ll M^2_{H_2}, \theta \ll 1$)
\begin{equation}
\Gamma (H_2 \rightarrow Z_{BL} Z_{BL}) \approx \frac{g^2_{BL} M^3_{H_2}}{8 \pi M^2_{Z_{BL}}}.
\end{equation}
Now, increase in $g_{BL}$ corresponds to smaller $M_{H_2}$ as evident by combining equations (\ref{eq:HiggsEigen}) and (\ref{eq:Zmass}) for a fixed $M_{Z_{BL}}$. Hence in general, enhancement of $g_{BL}$ does not always mean higher value of $\Gamma (H_2 \rightarrow Z_{BL} Z_{BL})$. However numerically, we find that for the chosen benchmarks of Fig. \ref{fig:FimpY}, even though $M_{H_2}$ decreases with increase in $g_{BL}$, the above decay width still increases by a factor of order one which enhances the yield of $Z_{BL}$ by some amount. Secondly, a lighter $H_2$ will have comparatively lesser Boltzmann suppression in its equilibrium number density. These two factors, with the latter being dominant, lead to the enhancement of $Z_{BL}$ (approximately by order of two), given other relevant parameters remain same. The production of $Z_{BL}$ from $H_1$ decay will be mixing suppressed due to smallness of $\lambda_3$. It is in fact kinematically forbidden for the chosen benchmark values of $Z_{BL}$ mass. For some epochs the abundance of $Z_{BL}$ remains constant (denoted by the plateau region) and then gets reduced to zero again due to subsequent decays of $Z_{BL}$ into $N_1$ as well as other lighter particles.
\begin{figure}[h]
\noindent \begin{centering}
\includegraphics[scale=0.35]{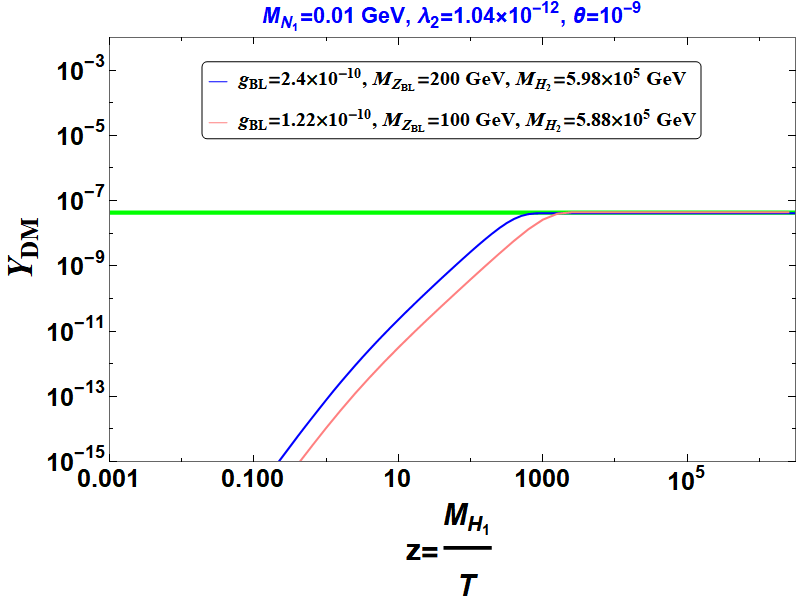}
\par\end{centering}
\caption{Evolution of comoving number densities for the DM $N_1$ as function of temperature for two different sets of $(g_{BL},M_{Z_{BL}})$ as tabulated in table \ref{tab:FimpPara}. Note that, the two set of reference points used here gives correct relic abundance (green region) in the present universe.}
\label{fig:FimpRelic}
\end{figure}

Similar features can be observed in right panel of Fig. \ref{fig:FimpY} where the evolution of $N_1$ abundance is shown using the same choice of parameters as in left panel. The $N_1$ abundance begins from vanishingly small value and gets enhanced due to non-thermal production from $Z_{BL}$ and $H_2$ decays and finally gets saturated. We notice that larger $g_{BL}$ value leads to larger final abundance of the DM due to both the enhanced abundance of $Z_{BL}$ (as earlier mentioned) as well as larger partial decay width of $Z_{BL}$ into DM pairs. It is also relevant to mention here that in our working regime $M_{N_1}\ll M_{Z_{BL}}$, the associated Yukawa coupling ($Y_{N_1}$) with $H_2$ is suppressed compared to $g_{BL}$ and hence direct production of DM is primarily dominated from tree level $Z_{BL}$ decay.


 Once the freeze-in abundance of DM that is $Y_{\rm DM}$ saturates, one can obtain the present relic abundance using the following expression:
\begin{align}
\Omega_{\rm DM} h^2=2.755\times 10^{8} \left(\frac{M_{N_1}}{\rm GeV}\right)Y_{\rm DM}^{\rm present}.\label{eq:relicFIMP}
\end{align}
Here $\Omega_{\rm DM}=\frac{\rho_{\rm DM}}{\rho_c}$, where $\rho_{\rm DM}$ is the DM energy density and $\rho_c=\frac{3\mathcal{H}_0^2}{8\pi G_N}$ is the critical energy density of the universe, with $G_N$ being Newton's gravitational constant and $\mathcal{H}_0\equiv 100\: h~\text{km\: s}^{-1}\:\text{Mpc}^{-1}$ is the present-day Hubble expansion rate.

Using the above equation \eqref{eq:relicFIMP}, we now find some benchmark parameters of our model which satisfy the correct DM abundance in the present universe. In Fig. \ref{fig:FimpRelic}, we have shown the DM yield evolutions for two set of parameters that matches with the observed relic bound (green shaded region) at $z\rightarrow \infty$. In table \ref{tab:FimpPara} we list the numerical values of the parameters used in Fig. \ref{fig:FimpRelic}. As mentioned earlier, for such benchmark values of parameters the contribution of $2\rightarrow 2$ scattering processes to DM production in the present analysis remains sub-dominant or negligible. It should be noted that while the required FIMP DM relic abundance can be successfully generated in this model, the corresponding parameter space leads to decoupling of $B-L$ gauge sector from inflationary dynamics leading to a usual quartic plus non-minimal inflation \cite{Okada:2010jf}.
\begin{table}
\begin{center}
  \begin{tabular}{ | l | l | l | l | l | l | l | }
    \hline
    $g_{BL}$ & $M_{Z_{BL}}$ & $M_{H_2}$ & $\lambda_2$ & $Y_{N_1}$ & $Y_{N_{2}}(Y_{N_3})$&  $\sin\theta$  \\ \hline
    $2.4\times 10^{-10}$ & $200$ GeV & $5.98\times 10^5$ GeV & $1.04\times10^{-12}$ & $1.7\times10^{-14}$ & $10^{-6}(3\times 10^{-6})$ & $10^{-9}$\\ \hline
    $1.22\times 10^{-10}$ & 100 GeV & $5.88\times 10^5$ GeV & $1.04\times10^{-12}$ & $1.7\times10^{-14}$ & $10^{-6}(3\times 10^{-6})$ & $10^{-9}$\\ 
    \hline
  \end{tabular}
\end{center}
\caption{Two sets of parameters which can account for correct relic abundance for the FIMP case taking $\xi=0.01$ from the inflationary dynamics, considering $H_2$ can be produced
thermally ($M_{H_2} < T_R$).}
\label{tab:FimpPara}
\end{table}

So far, the analysis on non thermal production of dark matter is performed by assuming $H_2$ in thermal equilibrium with the SM bath. This is possible when $M_{H_2} < T_R$ and $H_2$ has sizeable couplings with other particles in the bath. However, it is also possible that $M_{H_2}$ remains larger compared to the reheat temperature $M_{H_2}>T_R$ and hence the inflaton remains out of equilibrium afterwards (see blue coloured region of Fig. \ref{fig:reheatII}). In such a case, the production of $Z_{BL}$ and $N_1$ will not be possible like the way it was discussed before. Since SM Higgs mixing with $H_2$ is also very small, it is not possible to generate correct FIMP abundance. While interactions by virtue of gauge coupling and Yukawa coupling with $H_2$ are insufficient to produce correct FIMP abundance, one can turn to Yukawa couplings with ordinary leptons which are present in thermal bath for most of the epochs. However one has to get rid of the $Z_2$ symmetry in order to introduce such Yukawa couplings through SM Higgs. We briefly discuss this possibility in the remainder of this section.

Once the $Z_2$ symmetry is discarded, one can have new non-diagonal terms in the RHN mass matrix. However, for simplicity we continue to choose a diagonal RHN mass matrix or the corresponding Yukawa coupling matrix $Y_N$. The newly introduced Yukawa couplings of $N_1$ to SM leptons can be written as
\begin{align}
-\mathcal{L_Y}\supset \sum_{\alpha =e, \mu, \tau}(Y_{D})_{1\alpha }\overline{l_L}^\alpha\tilde{H}N_{R_1},
\end{align}
This will generate mixing of $N_1$ with active neutrinos once the electroweak symmetry is broken. Using Casas-Ibarra parametrisation of equation \eqref{casas} and using the form of complex orthogonal matrix given in equation \eqref{Rmatrix}, the Yukawa coupling of $N_1$ with leptons can be expressed as
\begin{align}
(Y_D)_{ 1 \alpha}^T=\frac{\sqrt{2}}{v}\begin{pmatrix}
        0.146 \sqrt{m_3}\sqrt{M_{N_1}}\sin \gamma^\prime\\
        0.648 \sqrt{m_3} \sqrt{M_{N_1}} \sin \gamma^\prime\\
        0.746 \sqrt{m_3} \sqrt{M_{N_1}}\sin \gamma^\prime
        \end{pmatrix}\label{eq:Yu1}
\end{align}
where $\gamma^{\prime}$ is a complex angle and $m_3$ the heaviest active neutrino mass with normal ordering. In deriving this, we fix Dirac CP phase to be zero\footnote{Although recent experimental results hint towards a non-vanishing leptonic CP phase \cite{Abe:2019vii}, it does not affect our analysis significantly.} and also considered the lightest active neutrino as massless. The requirement of the lightest active neutrino mass to be vanishingly small arises due to tiny Yukawa couplings of $N_1$ to leptons for being a FIMP DM. We define the mixing of sterile $N_1$ with $i^{\rm th}$ active neutrino by:
\begin{align}
 \tan{\delta_i}=-\frac{\sqrt{2}~(Y_D)_{1i}v}{M_{N_1}}.
\end{align}
For simplicity, we redefine $\delta_1=\delta$ and the relation between $\delta$ and $\delta_{2,3}$ can be easily found using equation (\ref{eq:Yu1}). Owing to this tiny but non-zero mixing, $N_1$ can now interact with SM bath directly without relying upon $Z_{BL}$ or $H_2$ mediation considered earlier in $Z_2$ symmetric scenario. For example, $W^\pm$ boson can directly decay to $N_1$ through $W^\pm\rightarrow N_1 \alpha^\pm, \alpha \equiv (e, \mu, \tau)$ if kinematically allowed. The contribution from annihilation processes continues to be sub-dominant like before. The evolution of DM comoving number density is governed by
\begin{align}
 \frac{dY_{\rm DM}}{dz}=&\frac{2 M_P}{1.66 M_{H_{1}}^2} \frac{z\sqrt{g_*(z)}}{g_{*s}(z)}\Bigg(\langle\Gamma_{H_{1}\rightarrow \nu_{\alpha} N_1}\rangle(Y^{\rm eq})+\langle\Gamma_{W^\pm\rightarrow e^\pm N_1 }\rangle (Y^{\rm eq})\Bigg),
\end{align}
where we have considered only the most dominant decay modes and completely ignored the annihilation processes which are sub-dominant. Decay channels with more than one $N_1$ in final state will be suppressed due to higher powers of tiny mixing $\delta$. Once we obtain $Y_{\rm DM}$, it is simple to compute the relic density of the DM using equation (\ref{eq:relicFIMP}) discussed earlier. It turns out that the DM relic abundance is primarily determined by the decay of $W^\pm$ (with other RHNs very heavy compared to DM) which further depends crucially on the mixing parameter $\delta$. 
\begin{figure}[h]
\noindent \begin{centering}
\includegraphics[height=7cm,width=10cm]{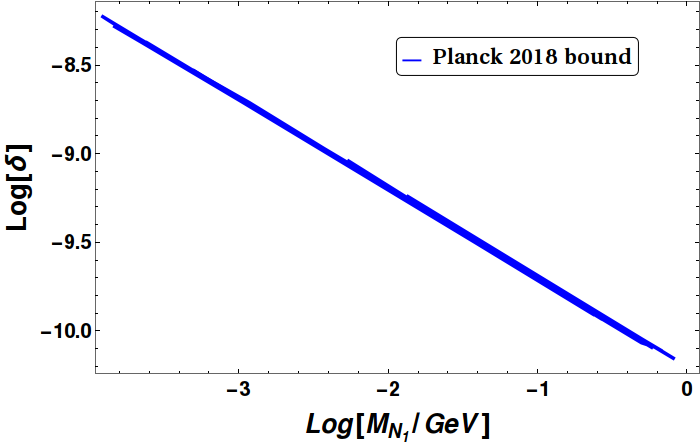}
\par\end{centering}
\caption{Contour for observed relic abundance in $\delta-M_{N_1}$ plane considering $H_2$ to be out of equilibrium and DM production from tree level decay of $W^\pm$ boson.}
\label{fig:Fimp2Relic}
\end{figure}
In Fig. \ref{fig:Fimp2Relic}, we show the contour for the observed relic abundance in $M_{N_1}-\delta$ plane.  The figure shows the dependence of relic abundance on both DM mass the mixing $\delta$ with lower $M_{N_1}$ requiring larger $\delta$, as expected. The magnitude of $\delta$ ($Y_{1e}$) is required to be extremely small to generate correct order of DM relic abundance. Such a tiny Yukawa element can be obtained by suitable value of free parameter $\gamma^\prime$ in equation (\ref{casas}). While generating figure \ref{fig:Fimp2Relic}, we assume  $M_{Z_{BL}}=10^4$ GeV, $M_{N_2}=10^9$ GeV,  $M_{N_3}=3\times 10^9$ GeV and $\lambda_2=4.35\times 10^{-10}$ (corresponding to $\xi=1$) with $\lambda_3=10^{-10}$, $g_{BL}=10^{-12}$ at inflationary energy scale. For these set of values, $H_2$ remains out of equilibrium after reheating. We have also confirmed that the contour for the observed relic abundance remains more or less same with different orders of of $\lambda_2,~\lambda_3$ and $g_{BL}$ provided $\lambda_3\lesssim\lambda_2$ and $M_{H_2}>T_R$. This is expected since here DM gets produced from $W$ boson decay which stays in thermal equilibrium.

It is to be noted that, unlike the WIMP scenario, we are not performing a complete scan of parameter space for FIMP which can be found elsewhere. We have considered two possibilities based on inflaton mass being smaller or larger compared to reheat temperature and showed that required FIMP DM abundance can be successfully produced in both the scenarios. In the case where inflaton mass is larger compared to reheat temperature so that it is not present in the thermal bath afterwards, we find that the correct FIMP abundance can be produced only when we discard the $Z_2$ stabilising symmetry of DM and allow for more possibilities of its production from SM bath to open up. It is relevant to note here that such removal of $Z_2$ symmetry could produce extra relic through Dodelson-Widrow mechanism \cite{Dodelson:1993je}. However, considering the smallness of $\delta$ we have obtained to satisfy the observed relic limit, this effect is expected to be negligible. On the other hand, such long-lived dark matter can have very interesting consequences at indirect detection experiments, which have been summarised in the review article \cite{Adhikari:2016bei}.
\section{Leptogenesis}\label{sec:baryo}
In this section, we briefly discuss the possibilities of generating the observed baryon asymmetry of the universe through leptogenesis. Since the lightest right handed neutrino is our DM candidate, the required lepton asymmetry can be generated only by the out of equilibrium decays of heavier right handed neutrinos $N_{2,3}$. Usually, in such type I seesaw framework, the requirement of producing the correct lepton asymmetry pushes the scale of right handed neutrinos to a very high scale $M >10^9$ GeV, known as the Davidson-Ibarra bound \cite{Davidson:2002qv} of high scale or vanilla leptogenesis. For right handed neutrino masses lower than this, say around TeV scale, it is still possible to generate correct lepton asymmetry by resorting to a resonant enhancement of the CP-asymmetry with a quasi-degenerate right handed neutrino spectrum~\cite{Pilaftsis:2003gt, Dev:2017wwc}, known as resonant leptogenesis. In both vanilla as well as resonant leptogenesis, it is assumed that right handed neutrinos were produced thermally in the early universe along with other SM particles. For earlier works on thermal leptogenesis in gauged $B-L$ model, please refer to \cite{Iso:2010mv, Heeck:2016oda, Dev:2017xry} and references therein. Due to the presence of gauge interactions of right handed neutrinos in this model, there exist additional washout processes erasing the created asymmetry which leads to tight constraints on such $B-L$ gauge sectors, specially for low scale leptogenesis. Since we find thermal DM to be disfavoured in our model, we therefore do not discuss thermal leptogenesis any further. Also, thermal leptogenesis is not affected much by inflationary dynamics at high scale. It is of course possible to realise thermal leptogenesis and non-thermal DM in this model, but we focus mainly on non-thermal leptogenesis due to its connection to inflation as well as reheat temperature as discussed below. In fact, thermal vanilla leptogenesis is not possible in our setup as the predicted values of reheat temperature (for $g_{BL},\Sigma_N^4\ll\lambda_2^2$) discussed earlier (see Fig. \ref{fig:reheatII}) falls below the Davidson-Ibarra limit on scale of such leptogenesis. This motivates us to discuss non-thermal leptogenesis in this section.

The scenario of non-thermal leptogenesis \cite{Lazarides:1991wu, Giudice:1999fb, Asaka:1999yd, Asaka:1999jb, Fujii:2002jw, Pascoli:2003rq, Asaka:2002zu, Panotopoulos:2006wj, HahnWoernle:2008pq}  arises when the reheat temperature after inflation is lower than the masses of right handed neutrinos. Thus, although the right handed neutrinos can be produced due to the decay of inflaton, they cannot reach thermal equilibrium with the SM particles due to insufficient reheat temperature. The non-equilibrium abundance of right handed neutrinos will be purely decided by their couplings to inflaton which will affect the final CP asymmetry generated by subsequent decays of right handed neutrinos. Since inflaton also has to decay into other SM bath particles reproducing a radiation dominated universe, one has to solve coupled Boltzmann equations involving inflaton, right handed neutrinos and SM radiation. However, for simplicity, we assume that the decay width of $N_{2,3}$'s ($\Gamma_{N_{2,3}}$) to be larger than that of the inflaton ($\Gamma_{H_2}$) so that decays of $N_{2,3}$ to SM particles can be instantaneous \cite{Asaka:2002zu}. This allows us to retain the same reheating description (from inflaton decay only) discussed earlier. Thus, the right handed neutrinos produced from inflaton decay turns non-relativistic and decays to SM leptons and Higgs instantaneously. The CP asymmetry generated by $N_i$ decays, following the notations of \cite{Pascoli:2003rq}, can be formulated as
  \begin{align} 
   \epsilon_A&= \sum_{i=2}^3\frac{\Gamma(N_i\rightarrow H+l_L)-\Gamma(N_i\rightarrow H^\dagger+\overline{l_L})}{\Gamma(N_i\rightarrow H+l_L)+\Gamma(N_i\rightarrow H^\dagger+\overline{l_L})}=\epsilon_{A}^{2}+\epsilon_{A}^{3}\\
   &=\frac{1}{8\pi}\frac{{\rm Im}\Big[\Big(Y_DY_D^\dagger\Big)_{23}\Big]^2}{\Big(Y_DY_D^\dagger\Big)_{22}}\mathcal{G}\Big(\frac{M_{N_3}}{M_{N_2}}\Big)+\frac{1}{8\pi}\frac{{\rm Im}\Big[\Big(Y_DY_D^\dagger\Big)_{32}\Big]^2}{\Big(Y_DY_D^\dagger\Big)_{33}}\mathcal{G}\Big(\frac{M_{N_2}}{M_{N_3}}\Big),\label{eq:assym}
  \end{align}
where the first and second terms in equation (\ref{eq:assym}) are the individual contributions of $N_2$ and $N_3$ respectively. The loop function $\mathcal{G}(x)$ containing both self-energy and vertex corrections is defined as
\begin{align}
 \mathcal{G}(x)=-x\Bigg[\frac{2}{x^2-1}+{\rm ln}\Big(1+\frac{1}{x^2}\Big)\Bigg].
\end{align}
Once the CP asymmetry parameter is calculated, the comoving lepton asymmetry (ratio of excess of leptons over antileptons and entropy) can be calculated as 
\begin{align}
 \frac{n_L}{s}=\epsilon_A^2{\rm Br}_2\frac{3 T_R}{2 M_{H_2}}+\epsilon_A^3{\rm Br}_3\frac{3 T_R}{2 M_{H_2}},
\end{align}
where Br$_i$ represents the branching ratio of the inflaton decay to $N_i$. Finally, the baryon asymmetry generated through the standard sphaleron conversion processes is given by
\begin{align}
Y_B= \frac{n_B-n_{\bar{B}}}{s}=-\frac{28}{79}\frac{n_L}{s}.
\end{align}
\begin{figure}[h] 
\noindent \begin{centering}
\includegraphics[height=10cm,width=13.5cm]{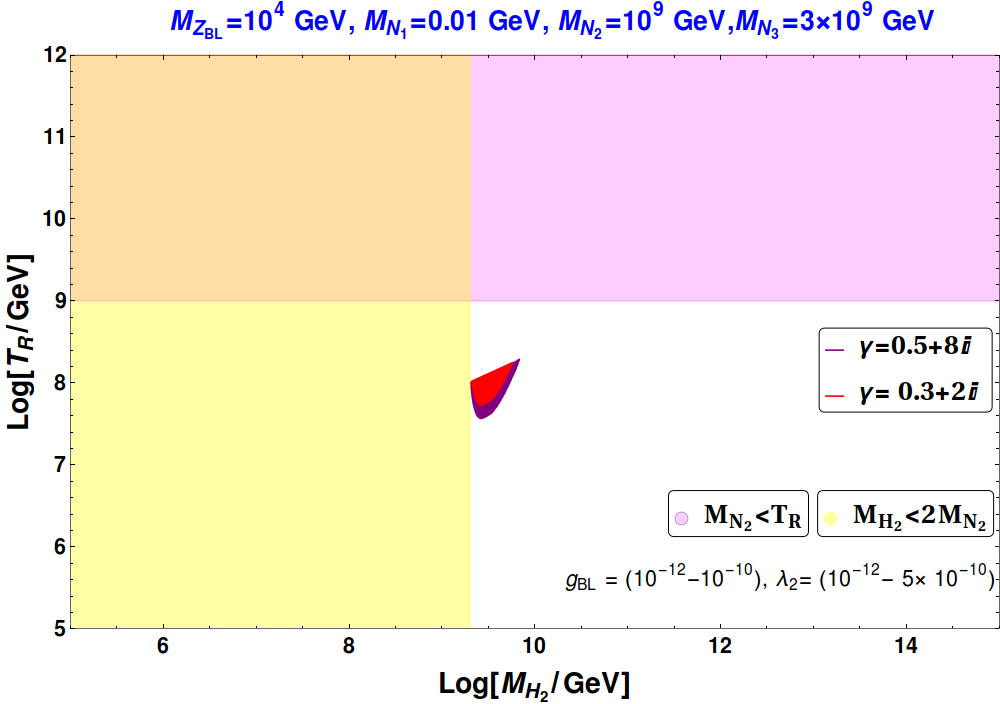}
\end{centering}
\caption{Region allowed by the observed baryon asymmetry in $M_{H_2}-T_R$ plane by varying $g_{BL}$, $\lambda_2$ and angle $\gamma$ considering $M_{N_2}=10^9$ GeV. We also include the essential conditions to realize the non thermal leptogenesis such as $M_{N_2}>T_R$, $M_{H_2}>2 M_{N_2}$ in the figure.}
\label{fig:Bar1}  
\end{figure} 
We have used the Casas-Ibarra parametrisation of $Y_D$ as given by equation (\ref{casas}). Since lepton asymmetry gets generated from $N_2$ and $N_3$ decays, the complex angle $\gamma$ in equation (\ref{Rmatrix}) is an important parameter to be tuned appropriately. Note that there is not much freedom to choose $\gamma^{\prime}$ as it appears in FIMP DM coupling discussed earlier. We consider it to be vanishingly small for leptogenesis discussions. As in the preceding analysis, here also we consider $M_{N_3}=3\times M_{N_2}$. Thus it is expected that $N_2$ will dominantly contribute to the baryon asymmetry.

It is to be noted that in the present scenario the inflaton has several other decay modes, in addition to its decay into RHNs. Thus it is difficult to generate the observed amount of baryon asymmetry where the inflaton decays to RHNs are subdominant or ${\rm Br}_{\phi\rightarrow N_{2,3}N_{2,3}}\ll 1$. So, one needs to find the parameter space where the branching ratio of inflaton to RHNs as well as the CP asymmetry from RHN decay can be large enough to satisfy the requirement of baryon asymmetry. The decay widths of RHNs $N_2$ and $N_3$ into SM leptons and Higgs depend
on the strength of Yukawa couplings as defined in 
equation (\ref{casas}). Below we provide the structure of $Y_{D_{2i}}$ and $Y_{D_{3i}}$ (see equation (\ref{eq:Yu1}) for $Y_{D_{1i}}$) where we have considered best fit values of light neutrino mass parameters with vanishing Dirac CP phase\footnote{Even if we take non-vanishing Dirac CP phase, as suggested by recent experiment \cite{Abe:2019vii}, it does not appear in the calculation of lepton asymmetry in unflavoured regime.} and vanishing lightest active neutrino mass (normal ordering).
\begin{align}
 Y_{D_{2,i}}^T=\frac{\sqrt{2}}{v}
              \begin{pmatrix}
              0.56 \sqrt{m_2}\sqrt{M_{N_2}}\cos\gamma+0.146 \sqrt{m_3}\sqrt{M_{N_3}}\cos\gamma^\prime \sin\gamma\\
              0.56 \sqrt{m_2}\sqrt{M_{N_2}}\cos\gamma+0.648 \sqrt{m_3}\sqrt{M_{N_3}}\cos\gamma^\prime \sin\gamma\\
              -0.60 \sqrt{m_2}\sqrt{M_{N_2}}\cos\gamma+0.746 \sqrt{m_3}\sqrt{M_{N_3}}\cos\gamma^\prime \sin\gamma
             \end{pmatrix}
\end{align}

\begin{align}
 Y_{D_{3,i}}^T=\frac{\sqrt{2}}{v}
              \begin{pmatrix}
              0.146\sqrt{m_3}\sqrt{M_{N_3}}\cos\gamma\cos\gamma^\prime-0.56\sqrt{m_2}\sqrt{M_{N_3}}\sin\gamma\\
              0.648\sqrt{m_3}\sqrt{M_{N_3}}\cos\gamma\cos\gamma^\prime-0.56\sqrt{m_2}\sqrt{M_{N_3}}\sin\gamma\\
              0.746\sqrt{m_3}\sqrt{M_{N_3}}\cos\gamma\cos\gamma^\prime+0.60\sqrt{m_2}\sqrt{M_{N_3}}\sin\gamma
              \end{pmatrix}
\end{align}

\begin{figure}[h] 
\noindent \begin{centering}
\includegraphics[height=10cm,width=13.5cm]{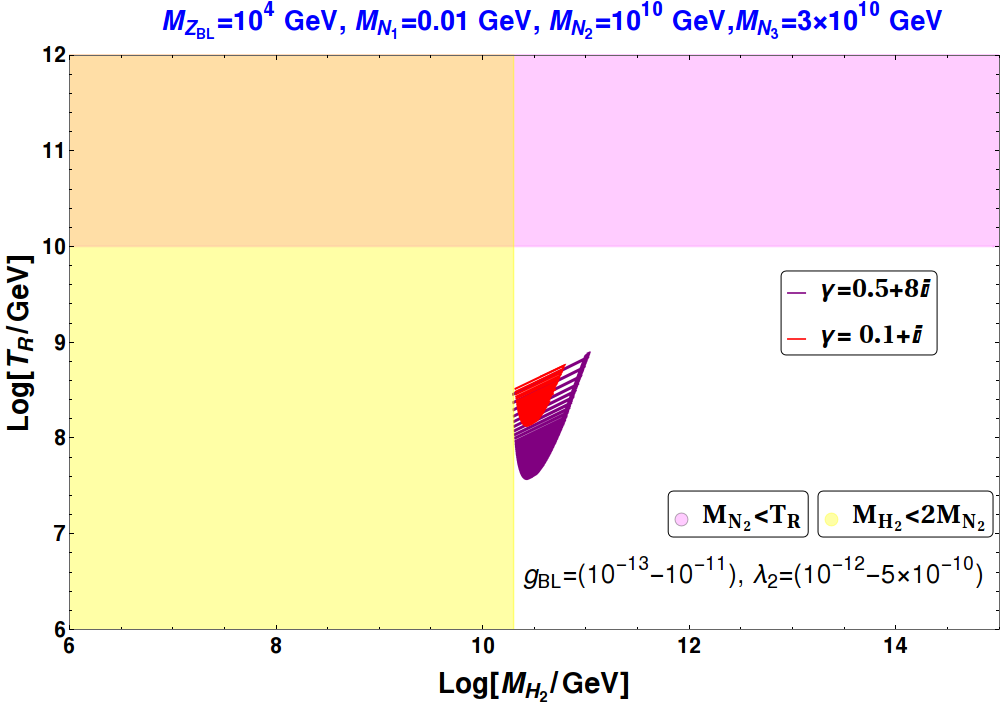}
\end{centering}
\caption{Region allowed by the observed baryon asymmetry in $M_{H_2}-T_R$ plane by varying $g_{BL}$, $\lambda_2$ and angle $\gamma$ considering $M_{N_2}=10^{10}$ GeV. We also include the essential conditions to realise the non thermal leptogenesis such as $M_{N_2}>T_R$, $M_{H_2}>2 M_{N_2}$ in the figure.}
\label{fig:Bar2}  
\end{figure}

In Fig. \ref{fig:Bar1}, we show the allowed region which satisfies the bound on $Y_B$ in $M_{H_2}-T_R$ plane for two different sets of complex angle $\gamma$ considering $M_{N_2}=10^9$ GeV. We vary $g_{BL}$ and $\lambda_2$ in specified ranges mentioned in the figure. The regions labelled as $M_{N_2}<T_R$ and $M_{H_2}<2 M_{N_2}$ in magenta and yellow colours respectively are outside the regime of non-thermal leptogenesis discussed here. Similar plot is shown in Fig. \ref{fig:Bar2} considering slightly higher scale of leptogenesis ($M_{N_2}=10^{10}$ GeV) where the allowed region gets enhanced, as expected. In preparing both the figures we have taken $\lambda_3\sim \mathcal{O}(10^{-15})$, such that the Br$_{\phi\rightarrow N_{2,3}N_{2,3}}$ does not turn very small due to other decay modes of inflaton which depend upon $\lambda_3$ or scalar mixing. We have also confirmed that corresponding to our choices of $\gamma$, the condition $\Gamma_{N_{2,3}}\gg \Gamma_{H_2}$ is satisfied, a requirement for validating the simplistic approach adopted here.

\section{Conclusion}\label{sec:conclude}
 To summarise, we have studied the very popular gauged $B-L$ extension of the standard model by restricting ourselves to the minimal possible framework from the requirement of triangle anomaly cancellation, desired gauge symmetry breaking and origin of light neutrino mass. We particularly focus on the possibility of singlet scalar field responsible for breaking $B-L$ gauge symmetry spontaneously to also drive successful inflation in agreement with Planck 2018 data and its implications for dark matter and leptogenesis. While the lightest right handed neutrino is considered to be the DM candidate, the heavier two right handed neutrinos generate light neutrino masses through type I seesaw mechanism and also generate the required lepton asymmetry via their out of equilibrium decays. We first show that the requirement of successful inflationary phase tightly constrains the scalar and gauge sector couplings of the model. To be more precise, the requirement of stability of the inflationary potential puts an upper bound on $B-L$ gauge coupling along with inflaton couplings to SM Higgs as well as right handed neutrinos. Since WIMP type DM in this model primarily interacts with the SM particles via $B-L$ gauge or singlet scalar (via its mixing with SM Higgs), the bounds derived from inflation on couplings and masses involved in these portals make WIMP annihilations inefficient. The parameter space where WIMP abundance satisfies the Planck 2018 data on DM abundance along with inflationary requirements, gets ruled out by LHC data on dilepton searches. This led to our first main conclusion that thermal DM is disfavoured in such scenario. We then considered the possibility of non-thermal DM by considering two different broad scenarios related to the interplay of inflaton mass and reheat temperature. We show that in both the scenarios correct FIMP abundance can be produced. We find that for a scenario where inflaton is not part of the thermal bath after reheating, the required FIMP relic can be produced only if it is allowed to couple to SM leptons opening up several production channels from the SM bath. Such a scenario does not require any additional $Z_2$ symmetry considered for stabilising WIMP type DM and also have interesting consequences for indirect detection experiments due to possible decays into photons ranging from X-ray to gamma rays.
 
We then briefly discuss the possibility of leptogenesis by focusing primarily on non-thermal leptogenesis which is very much sensitive to the details of inflation. While resonant leptogenesis is still a viable option, thermal vanilla leptogenesis is not possible due to low reheat temperature predicted in our scenario. We find that inflationary requirements tightly constrain the scenario of non-thermal leptogenesis, precisely due to the same reason behind constraining or disfavouring WIMP type DM mentioned earlier. We show the possibility of producing observed baryon asymmetry from non-thermal leptogenesis for benchmark choices of some parameters while varying others and also show that the same parameters are also consistent with successful inflation, stability of inflaton potential, FIMP DM abundance, neutrino mass apart from other experimental limits. Since the model is very minimal, it remains very predictive, specially when the requirements of correct neutrino mass, DM abundance, baryon asymmetry along with successful inflation are to be met with. Future data from all these frontiers should be able to restrict the model parameters to even stricter ranges while ruling out some of the possibilities. 

Before we end, let us briefly comment on the fate of electroweak vacuum in view of our proposed inflationary scenario. During inflation, quantum fluctuations of the Higgs field are developed with amplitude proportional to the Hubble parameter during inflation $\sim H_{\rm Inf}$. This could be dangerous since the electroweak vacuum in the SM is metastable \cite{Isidori:2001bm,Buttazzo:2013uya,Tang:2013bz,Anchordoqui:2012fq} and it is expected to remain same in our framework as well due to the small mixing angle between SM Higgs and singlet scalar. Usually in large scale inflation models, $H_{\rm Inf}$ turns bigger than the instability scale of the SM Higgs vacuum ($\sim 10^9$ GeV \cite{Isidori:2001bm}) and therefore, during inflation, the Higgs field can cross the potential barrier towards the unbounded part \cite{Kobakhidze:2013tn}. This serious drawback of large scale inflation model can be easily avoided by introduction of inflaton-Higgs quartic coupling. In that case, due to super-Planckian value of inflaton field, the Higgs field acquires inflaton dependent effective mass during inflation which becomes larger than the Hubble scale. Then, the quantum fluctuations of the Higgs field can be ignored. This holds in our analysis as well. However, some studies \cite{Herranen:2015ima,Kohri:2016wof,Ema:2016kpf} have shown that the stability of the electroweak vacuum is essential even after inflation as oscillation phase of the inflaton could trigger resonant enhancement of the Higgs fluctuations. Addressing the post inflationary Higgs instability is beyond the scope of our present work and introduction of additional degree of freedom in form of a scalar field may be useful to ensure this (see Ref. \cite{Ema:2016kpf}, for example). We leave such studies with next to minimal extension of the present model to future works.

\acknowledgements

DB acknowledges the support from Early Career Research Award from DST-SERB, Government of India (reference number: ECR/2017/001873). SJD would like to thank Dibyendu Nanda and Devabrat Mahanta for some productive discussions. AKS is thankful to Rome Samanta for some useful discussions during WHEPP 2019. AKS also acknowledges PRL for providing postdoctoral research fellowship.

\appendix

\section{RGE Equations}
\label{appen1}
Here we present the complete set of RGEs at one loop level for the minimal B-L model:
\begin{align}
&\beta_{\lambda_{1}}=24\lambda_{1}^{2}+\lambda_{3}^{2}-6Y_{D}^{4}+\frac{9}{8}g_1^{4}+\frac{3}{8}g_{2}^{4}+\frac{3}{4}g_1^{2}g_{2}^{2}+12\lambda_{1}Y_{D}^{2}-9\lambda_{1}g_1^{2}-3\lambda_{1}g_{2}^{2}\\
&\beta _{\lambda_{3}} = \lambda _3\Bigg( 3\lambda _1+2\lambda _2+\lambda _3+\frac{3}{2}Y_D^2-\frac{9}{8}g_1^2-\frac{3}{8}g_2^2+\Sigma_N^2-6g_{BL}^2\Bigg)\\
&\beta_{g_{s}}=-7g_{s}^3\\
&\beta_{g_1}=-\frac{19}{6}g_1^3\\
&\beta_{g_{2}}=\frac{41}{6}g_{2}^3\\
&\beta_{Y_{D}}=Y_{D}\Bigg(\frac{9}{2}Y_{D}^2-8g_{s}^2-\frac{9}{4}g_1^2-\frac{17}{12}g_{2}^2-\frac{2}{3}g_{BL}^2\Bigg).
\end{align}
where $g_s$, $g_1$ and $g_2$ represent the $SU(3)_C$, $SU(2)_L$ and $U(1)_Y$ gauge couplings respectively.


\providecommand{\href}[2]{#2}\begingroup\raggedright\endgroup

\end{document}